\documentstyle[preprint,aps]{revtex}
\tightenlines
\begin{document}

\begin{center}
\begin{large}
{\bf Orbital Polarization, Surface Enhancement and Quantum
Confinement in Nano-cluster Magnetism}
\end{large}

\vspace*{0.5cm}

Xiangang Wan$^{1}$, Lei Zhou$^{2}$, Jinming Dong$^{1}$, T.K.
Lee$^{3}$ and Ding-sheng Wang$^{4,5}$

\vspace{0.5cm}

$^{1}$National Laboratory of Solid State Microstructures and
Department of Physics, Nanjing University, Nanjing 210093, China

$^{2}$ Department of Physics, The Hong Kong University of Science and
Technology, Clear Water Bay, Kowloon, Hong Kong, China


$^{3}$Institute of Physics, Academia Sinica, Nankang, Taipei,
Taiwan 11529

$^{4}$Institute of Physics, Academia Sinica, Beijing 100080, China

$^{5}$Interdisciplinary Center of Theoretical Studies, Academia
Sinica, Beijing 100080, China

\end{center}
\begin{abstract}
Within a rather general tight-binding framework, we studied the magnetic properties
of Ni$_{n}$ clusters with $n=9$ through 60. In addition to usual hopping, exchange,
and spin-orbit coupling terms, our Hamiltonian also included orbital correlation and
valence orbital shift of surface atoms. We show that
orbital moment not only contributes appreciably to the total moment in this range of
cluster size, but also dominates the oscillation of total moment with respect to the
cluster size. Surface enhancement is found to occur not only for spin but, even
stronger, also for orbital moment. The magnitude of this
enhancement depends mainly on the coordination deficit of surface
atoms, well described by a simple interpolation. For
very small clusters ($n \leq 20$), quantum confinement of 4s
electrons has drastic effects on 3$d$ electron occupation, and thus greatly influences
both spin and orbital magnetic moments. With physically reasonable
parameters to account for orbital correlation and surface valence orbital shift,
our results are in good quantitative agreement with available experiments,
evidencing the construction of a unified theoretical framework
for nano-cluster magnetism.

\end{abstract}


\newpage
\section{Introduction }

Theoretical and experimental studies have demonstrated that
magnetism of materials is strongly affected by their
dimensionalities and sizes. This has been clearly illustrated in
multilayers and gas-phase clusters whose sizes can be reduced to
atomic level. The strong size dependence of magnetic moment opens
new possibilities to design materials with specific and tailored
properties. Recently, transition metal nano-clusters have
attracted considerable attention due to both theoretical and
practical interests. Using the Stern-Gerlach deflection technique,
Billas et al. \cite{billas1,billas} measured the magnetic moments
of Ni, Fe, and Co clusters ranging from about 25 to 700 atoms. It
was shown that small clusters usually possessed large magnetic
moments, e.g., about 1 $\mu_{B}$/atom for Ni$_{20-30}$. Using the
same technique, Apsel et al.\cite{apsel} performed more accurate
measurements for nickel clusters containing 5 to 740 atoms. In
addition to an overall decrease with increasing cluster size, they
found the magnetic moment to exhibit characteristic oscillations.
A pronounced sharp minimum was found at $n=13$, and other less
pronounced minima at $n=6$, 34 and 56 (see dashed lines in Fig.
1). Different from Billias et al., Apsel et al. found that the
approach to bulk magnetism was much slower. For example, the
magnetic moment was larger than 0.70 $\mu_{B}$/atom even up to
$n=500$. Very recently, Knickelbein\cite{knickelbein} again
measured the magnetic moments of small Ni clusters containing 7 to
25 atoms. While his results confirmed the pronounced sharp moment
minimum at Ni$_{13}$ (Fig. 1), the measured moment values were
about 0.2 $\mu_{B}$/atom smaller than those of Apsel et al for
most clusters. In general, all these experiments showed similar
size dependences of magnetic moment, in despite of some
quantitative differences which may come from either experimental
errors or the differences in the structure of the clusters.

In a theoretical discussion, Billias et al.\cite{billas} proposed
a simple magnetic shell model to account for the size dependences
of magnetic moments for Fe, Co and Ni clusters. They assumed their
clusters to be structureless and consist of several spherical
atomic shells. The magnetic moment of an atom on each shell
depends only on its distance to the cluster surface. While this
simple model yielded a correct decreasing trend of magnetic moment
versus size, it failed to reproduce the oscillations superimposed
on the overall decrease. It is well known that giant moments exist
on surface atoms, as revealed in early 1980s by either local spin
density functional (LSDA) calculations and later substantiated by
experiments (see, for example, a review by Freeman and
Wu\cite{freeman}). Table I lists some typical LSDA results for Ni
systems in bulk and film geometries from standard full potential
linearized augmented plan wave (FLAPW) calculations \cite{wimmer}.
For Ni systems, the spin moment per atom drops from 2 $\mu_B$ for
an isolated atom (Hund's rule), to 1.014 $\mu_B$ for an isolated
(100) monolayer (the coordination number $Z=4$) and 0.675 $\mu_B$
for a surface atom of (100) 5-layer slab ($Z=8$), and finally to
0.561 $\mu_B$ for a bulk crystal. Taking such $Z$ dependence into
account, Jensen and Bennenmann\cite{jensen} proposed a
coordination model for cluster magnetism, in which the magnetic
moment of an atom was assumed to depend solely on its number of
nearest neighbors $Z$ as in the film case. In the following
sections, this idea will be tested thoroughly. We found that, in
despite of the complicated local environments for atoms on a
cluster surface, this model still approximately holds for both
spin and orbital moments in cluster magnetism.

On the other hand, Fujima and Yamaguchi\cite{fujima1} considered
the effects of volume confinement for delocalized atomic {\it s}
electrons in clusters. Strong bonding, less strong bonding, or
anti-bonding states, depending on the global symmetry (denoted as
capital S, P, D ... ) of the linear combination coefficients in
the Tight-binding (TB) terminology, were formed by the delocalized
atomic {\it s} orbits. This volume confinement raises the energy
separations between shells with global S, P, D .. symmetry. Upon
changing the cluster size, whenever one of these global shells
goes through the Fermi level, which locates near the top of denser
{\it d} bands, the total number of $s$-electrons changes abruptly,
which in turn, changes the number of holes in the {\it d} bands.
This effect offers a possible explanation for the oscillatory size
dependence of magnetic moment, but Fujima and Yamaguchi provided
only a qualitative suggestion. In order to convincingly identify
the oscillations observed in experiments to this quantum
confinement effect, a more rigorous treatment of the $s$-$d$
coupling and cluster structures should be made.

Ideally, one expects that an {\it ab initio} calculation (for
example, in the LSDA framework) could automatically take the above
mentioned surface and quantum confinement effect into account,
since it can precisely determine the charge of electronic density
due to the spillover on the surface, and the redistributions of
the $s$-$d$ electrons. But calculated results reported by Reuse et
al. \cite{reuse} (open triangles in Fig.1) and by Reddy et
al.\cite{reddy} (open circles in Fig.1) were substantially smaller
by about 0.3-0.6 $\mu_B$/atom. And they failed to reproduce the
moment variations --- even the most pronounced sharp minimum at
Ni$_{13}$. Using spin-polarized discrete variational method,
Fujima and Yamaguchi also studied the magnetic properties of Ni
clusters\cite{fujima}. The calculated magnetic moment was 0.58
$\mu_{B}$/atom for Ni$_{19}$, agreeing well with Reddy et al, and
0.73 $\mu_{B}$/atom for Ni$_{55}$. In general, all calculations
based upon first principles showed poor agreements by
underestimating substantially the spin moment (Fig. 1).
Considering the successes of LSDA in bulk and surface magnetism,
where the calculated moments of 3$d$ metals and alloys are usually
in agreement with experiments within about 0.05 $\mu_B$/atom, the
failure in clusters is more or less out of initial expectation.

The tight-binding (TB) method has been used to deal with even
larger clusters. The results are, in general, similar to the first
principles ones. Using a model Hamiltonian which took into account
the electron spillover at cluster's surfaces, Weissmann's
group\cite{guevara} studied the size dependences of magnetic
moments for fcc and bcc clusters up to 177 atoms. However, similar
to the first principle results, their calculated moments were
again mostly smaller than the experimental ones. Alonso and his
colleagues also made a thorough study for Ni clusters
\cite{alonso,bouarab,rodriguez}. Using a special parameterization
of the orbital energies, they could get moment values much higher
than the LSDA results, and brought, especially in the small size
range, the moment values closer to the experimental one. However,
their results are far from satisfactory, and the agreement seems
somewhat an artifact in their parameterization scheme (see a
discussion in subsection IV.A).

We note that all those previous calculations mentioned above did
not consider the orbital magnetic moment. Through a general
argument, it was shown that orbital correlation has stronger
effects in low dimensional transition metal systems than in bulk
crystals by leading to orbital polarized ground
states\cite{zhou1}. This was demonstrated also by first principles
calculations for isolated or substrate supported linear chain
systems\cite{komelj}, and adatoms on metallic
substrates\cite{nonas}. Presumably, orbital interactions should be
much more sensitive to the details of the local structures.
Therefore, we expect that the environment dependence of orbital
moment, if not fully quenched, should be even more remarkable than
that of spin moment. Very recently, using a TB scheme which
includes orbital correlation and spin-orbit interactions,
Guirado-L\'{o}pez et al.\cite{guirado} studied the magnetic
moments for Ni clusters. Their results confirmed the importance of
orbital moments in Ni clusters. Since they neglected the low
symmetry clusters and considered only a few fcc or icosahedral
geometries, their paper did not show a rather complete size
dependence of the moment to compare with experiments. It is thus
desirable to investigate the magnetic moments for an as complete
as possible series of Ni clusters, with orbital correlation
included, to check whether the disagreement between experiments
and theoretical works is due to the neglect of orbital
contribution.

The present paper is organized in the following way:
the geometrical structures used will be discussed in Sec. II,
and a detailed description of our model Hamiltonian will be
given in Sec. III. Calculation results are discussed
in Sec. IV, in which the importance of
orbital contribution is highlighted.
We analyze the surface enhancement effect
in Sec. V, and the quantum confinement effect in Sec. VI.
Finally, Sec. VII will summarize the results of present study.

\section{Cluster Geometry}

Itinerant electron magnetism is sensitive to the atomic structure
of the system. The main factors to influence the magnetic
properties of small Ni clusters are i) low atomic coordination and
the relaxation of inter-atomic distances for surface atoms; ii)
quantum confinement of delocalized {\it s} electrons which
controls indirectly the number of holes in the {\it d} bands. The
first factor is basically a local geometrical effect, and the
latter one is global. Both factors may play important roles in
determining the electronic structure. A good geometrical structure
is therefore a crucial start in any attempt to interpret the
nano-cluster magnetism. Experimental determination of a cluster
geometry is difficult, since most clusters are too large for
spectroscopic probes but too small for diffraction probes.
Therefore, people usually employ theory to determine the cluster
structures and subsequently compare some calculated properties
with experiments. Unfortunately, the structures of most clusters
(say, over a few tens of atoms) have not been accurately
determined by {\it ab initio} calculations because of the enormous
computational works involved. Instead, geometrical structures of
nano-clusters are available mostly from calculations using
semi-empirical interatomic potential.

Geometries of Ni$_{n}$ with $n \leq$ 20 were obtained by
molecular-dynamics simulations with a semi-empirical manybody
potential\cite{lopez,gupta}, and it is found that for small Ni
clusters the structure is icosahedral\cite{bouarab}. This
structure has also been verified subsequently by
Montejano-Carrizales, Iniguez, Alonso and Lopez
(MIAL)\cite{montejano} using an embedded-atom method. For $n$
larger than 13, two types of icosahedral growth were proposed,
usually called as Mackey icosahedra (MIC) and polyicosahedra
(TIC)\cite{alonso,montejano}. According to the MIAL paper, the TIC
growth is favored for $14 \le n \le 27 $ and $57 \le n \le 67$,
but the MIC growth is favored for $28 \le n \le 56$. This
evolution of structural symmetry is shown in Fig. 2. This set of
structures, called as the MIAL structure below for convenience,
was the one that has been confronted in great details with
reactivity experiments\cite{alonso,montejano}.

For the purpose of easy comparison, we also use this MIAL
structure after the following brief analysis\cite{mial}. A simple
check is made by counting the change of the total nearest neighbor
bonds with increasing cluster size. In general, the bond number
should increase at least 3 through adding one atom which is put
simply on top of an atomic triangle without further relaxations to
bring in even more bonds. This is true for most MIAL clusters,
except Ni$_{21}$, Ni$_{29}$, Ni$_{33}$, Ni$_{37}$, Ni$_{40}$ and
Ni$_{59}$ where the increase of bond number is equal to or less
than 2 compared with Ni$_{20}$, Ni$_{28}$, Ni$_{32}$, Ni$_{36}$,
Ni$_{39}$ and Ni$_{58}$, respectively. So, we suspect that these
six MIAL structures are probably artifacts of the specific
potential, or due to insufficient structural optimizations. We
neglect these six members (Ni$_{21}$, Ni$_{29}$, Ni$_{33}$,
Ni$_{37}$, Ni$_{40}$ and Ni$_{59}$) from the MIAL list in our
following discussions, with the concern that possible structure
uncertainty may mislead our understanding of nano-cluster
magnetism.

Symmetry, shape, and the size evolution are important
for understanding the global behaviors of the delocalized electron states.
For a simple and straight characterization of these properties,
three principal axes and their corresponding rotational inertia
(marked as $I_1 \leq I_2 \leq I_3$) are calculated for the MIAL clusters.
Their ratios, $I_1 / I_3$ and $I_2 / I_3$, are plotted in Fig. 2.
At $n=13$, 26 and 28, and 55, $I_1 / I_3 = I_2 / I_3 = 1$,  indicating that
these structures are sphere-like.
In fact, Ni$_{13}$ and Ni$_{55}$ are completed icosahedral structures
centering on one atom. On the other hand,
the center of Ni$_{28}$ is on an empty tetrahedral site.
During the course of increasing size from one sphere-like cluster
to another one, i.e., from $n=13$ to 26 and 28, or from $n=28$ to 55,
Ni clusters first become prolate ellipsoid ---
with $I_3 > I_2 \approx I_1$, then change to oblate ellipsoid --- $I_3 \approx I_2 > I_1$.
This evolution of cluster shape leads to a drastic change of
the degeneracy of the quantumly confined delocalized $s$-states.
As the results, the behaviors of filling these states deviate greatly
from simple expectations based on a spherical jellium model
(see discussions below in section VI).

Two of the moment minima (i.e., $n=6$ and 13) on the size dependence of
magnetic moment (vertical lines in Fig. 2) correspond exactly to
the high symmetry spherical shapes, while the other one at $n=56$ is
pretty close to the high symmetry structure at $n=55$. This
hints to some relations between the moment minima and the cluster
symmetries. However, the relation is not exact, and there is an
exceptional minimum at $n=34$. Detailed theoretical calculations are necessary to
clarify the true relation between the moment minima and the cluster symmetry

\section{Model Hamiltonian and Methodology}

For transition metals, the TB Hamiltonian has been widely used
because the interaction matrix elements can be easily
parameterized to reproduce the {\it ab initio} results in very
good accuracy. Since the {\it s-p-d} hybridization plays a major
role in determining the magnetic properties of transition metals,
usually 3$d$, 4$s$ and 4$p$ valence orbits are included, and the
interaction matrix elements are parameterized by fitting the
results to the equilibrium bulk bands. These parameters are then
transferred to other non-periodic systems, such as disordered
solids, after proper scaling to account for the change of
interatomic distance. However, for the cluster problems, one
should be careful to include the surface effects, a crucial
ingredient in the cluster systems, in the TB Hamiltonian.

Following Hamiltonian is used in present paper to describe the Ni
clusters,
\begin{eqnarray}
H & = & [ \sum\limits_{iL\sigma}
          \epsilon_{il}^{0} c_{iL\sigma}^{\dag}c_{iL\sigma}
         +\sum\limits_{ij}\sum\limits_{LL'\sigma}
           t_{ij}^{LL'}c_{iL\sigma}^{\dag}c_{jL'\sigma}
        ]               \nonumber \\
  & + & H_{SOC}
    + H_{ee}            \nonumber \\
  & + & \sum\limits_{i'\sigma}
        [ \epsilon^0_{i's'} c_{i's'\sigma}^{\dag}c_{i's'\sigma}
        + t^{ss'}(Z_{i'})  ( c_{i's'\sigma}^{\dag}c_{i's \sigma}+c_{i's \sigma}^{\dag}c_{i's'\sigma}) ]
                        \nonumber \\
  & + & \sum\limits_{i'L\sigma}
        \Delta\epsilon_{i'} (n^{i'}_{s'})
           ( c_{i'L\sigma}^{\dag}c_{i'L\sigma}
            +c_{i's'\sigma}^{\dag}c_{i's'\sigma}  ).
\end{eqnarray}
Here c$_{iL\sigma}^{\dag}$ (c$_{iL\sigma}$)
is the operator of the creation (annihilation)
of an electron with spin $\sigma$ and
orbital quantum number $L=(l,m)$ at site {\it i}.
Subscripts $i'$ and $s'$ represent respectively the surface atoms
and an empty orbit attached to each surface atom,
as explained below.

The first term in Eq. (1) represents the usual bulk bands through
the bare orbital energies, $\epsilon_{il}^{0}$, and the
interatomic hopping integrals t$_{ij}^{LL'}$ between orbit $L$ at
atom $i$ and orbit $L'$ at its nearest neighboring atom {\it j}.
Values of these model parameters are taken from standard
reference\cite{papaconst}, which are obtained by fitting to LSDA
bands of ferromagnetic fcc bulk Ni, in the Slater-Koster
approximation and taking only the two-center hopping integrals.
Since the spin exchange will be considered in a separate
interaction term $H_{ee}$, the parameters here (see Table II) are
the average values for two spin channels listed in the reference
book\cite{papaconst}. As the interatomic distances in clusters are
certainly not uniform and differ more or less from the bulk
values, variation of the hopping integrals with the interatomic
distance $r_{ij}$ is assumed to follow a power law
$(r_{0}/r_{ij})^{l+l'+1}$, where $r_0$ is the bulk equilibrium
distance and $l$ and $l'$ are the orbital quantum numbers of the
two orbits involved in the hopping\cite{heine}.

The second term, H$_{SOC}$, is the spin-orbit coupling (SOC)
given by
\begin{equation}
H_{SOC}=\xi \sum\limits_{iL\sigma,L'\sigma'}
        \langle L\sigma\mid {\vec S_{i}}\cdot{\vec L_{i}}\mid
                L^{\prime}\sigma^{\prime}  \rangle
        c^{\dagger}_{iL \sigma}c_{iL^{\prime}\sigma^{\prime}},
\end{equation}
where the SOC strength is set as $\xi$=0.073 eV (Table II) for $d$ orbitals
according to Ref. \cite{bruno}.

The third term $H_{ee}$ is the intraatomic d-d electron-electron
interaction, including both Coulomb and exchange interaction, and
responsible for orbital and spin polarization. Since orbital
polarization plays an important role in low dimensional transition
metal systems, following our previous work\cite{zhou}, we adopt
the general concept of the LDA+U method\cite{liechtenstein} to
account for it. In a generalized Hartree-Fock approximation
including all possible pairings, the Hamiltonian reads
\begin{equation}
H_{ee} =
\sum\limits_{i,L\sigma,L^{\prime}\sigma^{\prime}}
V_{L\sigma,L^{\prime}\sigma^{\prime}}^{i}
c_{iL\sigma}^{\dag}c_{iL^{\prime}\sigma^{\prime}},
\end{equation}
where
\begin{eqnarray}
V_{L\sigma,L^{\prime}\sigma^{\prime}}^{i}
  & =
    & \sum\limits_{L_{2}L_{3}}
      [[U_{LL_{2}L^{\prime}L_{3}}
          n^{i}_{L_{2}{\overline \sigma},L_{3}{\overline\sigma}}
      +(U_{LL_{2}L^{\prime}L_{3}}-U_{LL_{2}L_{3}L^{\prime}})
          n^{i}_{L_{2}\sigma,L_{3}\sigma}]
          \delta_{\sigma\sigma^{\prime}}             \nonumber  \\
  & -
    & U_{LL_{2}L_{3}L^{\prime}}
        n^{i}_{L_{2}{\overline \sigma},L_{3}\sigma}
        \delta_{\overline\sigma \sigma^{\prime} } ]  \nonumber  \\
  & -
    & U(n^i-0.5) \delta_{LL^{\prime}} \delta_{\sigma\sigma^{\prime}}
    + J(n^{i\sigma}-0.5)\delta_{LL^{\prime}}\delta_{\sigma\sigma^{\prime}}.
\end{eqnarray}
Here
  $n^{i}_{L\sigma,L'\sigma'} =
    \langle C^{\dag}_{iL\sigma}C_{iL'\sigma'} \rangle$
is the single-site density matrix determined self-consistently,
$n^{i\sigma}=Tr[n^{i}_{L\sigma,L'\sigma}]$ is the electron number
of spin $\sigma$, $\bar \sigma$ means $-\sigma$, and
$n^i=\sum_{\sigma} n^{i\sigma}$ is the total electron number on
atom $i$. Being an extension of Eq. (5) of Ref.
\cite{liechtenstein}, present expression is rotationally invariant with
respect to both space and spin.

Matrix elements $U_{LL_{2}L^{\prime}L_{3}}$ satisfy
rotation summation relation as given by Eq.(6) of Ref. \cite{liechtenstein},
and can all be determined by two parameters,
namely, the average on-site Coulomb repulsion $U$ and the exchange $J$,
\begin{eqnarray}
U   &=  & \frac{1}{(2l+1)^2}\sum_{mm'}U_{mm'mm'}\\
U-J &=  &\frac{1}{2l(2l+1)}\sum_{mm'}(U_{mm'mm'}-U_{mm'm'm})
\end{eqnarray}

In this formalism, the Stoner parameter, which determines the
exchange splitting of the bulk bands\cite{zhou1},  is $I =
(2lJ+U)/(2l+1)$. To be compatible with the exchange splitting in
the LSDA bulk bands, $I = 1.12$ eV (Table II) is fixed in our
calculations. However, the correlation parameter $U$ is often set
adjustable, because its exact value may vary from system to system
even for atoms of the same element, say, about 5 eV in oxides, but
less than 3 eV in metallic system\cite{anisimov}. It is also
possible that, in the case of clusters, $U$ values could also be
different for clusters with different sizes, or even be different
for the center and boundary atoms in one single cluster. For
simplicity, such variations are neglected in present study. We
take $U=2.6$ eV in most calculations, and other values (1.8 and
3.2 eV, as in Table II) to check the influences of $U$ parameter
variation.

Although terms 1 through 3 in Hamiltonian (1) mainly depict the
physical behaviors exhibited in bulk systems, they could also
describe, at least partly, the physical effects due to the
coordination deficit of surface atoms. One example is the band
narrowing, which transfers the $s$ electrons to more than half
filled {\it d} bands leading to a surface moment reduction. This
is well known from the earliest TB studies back to 1970s before
the LSDA formalism. Later, however, LSDA revealed that giant
moments exist on 3$d$ transition metal surfaces, which was
subsequently confirmed by experiments \cite{freeman}. The physical
origin for this giant moment on surface is the energy shift of the
electronic orbits of surface atom, which lowers the electron
occupancy and increase the number of $d$-holes on the surface
atoms. This shift is due to the spillover of surface electrons and
the dipole layer thus formed on the surfaces, which has been
treated properly by standard all electron LSDA calculations for Ni
systems in the film geometry (see Table I for the core level
shift). However, to include properly this energy shift in a TB
study of the nano-cluster magnetism, where the surface is much
more complex than in the film geometry, is obviously a key
problem, but a toughest challenge.

To account for the electron spillover existing on cluster
surfaces, following Weissmann's group\cite{guevara}, we introduce
an extra orbit, $s'$, to each surface atom $i'$ (atom with
coordination number $Z<12$ for Ni clusters), and add the fourth
term to our Hamiltonian (1). We assume this orbit $s'$ to locate
in the vacuum outside the surface, has an {\it s} symmetry, an
energy $\epsilon_{s^{\prime}}^{0}$, and interact with {\it s}
orbit of the same surface atom by hopping integral $t^{ss'}$.
Electron occupation of this $s'$ orbit represents the spillover of
electrons from surface atoms to the vacuum. In the ideal film
geometry as treated by Weissmann's group, all surface atoms are
identical, and the orbital energy and hopping integrals of $s'$
orbit are constants. To account for the differences of
environments encountered by atoms on cluster surfaces, variations
should be introduced.

The smaller the local coordination, the larger the open space of this surface atom.
This change of open space volume could be accounted for approximately by putting
the hopping integral proportional to the square root
of the number of coordinate deficit,
\begin{equation}
  t^{ss'}(Z) =  V_{ss'\sigma} \sqrt{Z_{max}-Z}.
\end{equation}
Here $V _{ss'\sigma}$ is the hopping strength between $s$ and $s'$
orbits, $Z_{max}$ is approximately the maximum allowable number of
nearest neighbor spheres, and $Z_{max}-Z$ is thus the number of
empty volume around an atom with coordination number $Z$. Though
$Z_{max}=12$ seems to be a natural choice for Ni which exhibits
equilibrium fcc bulk structure, we prefer to take both
$V_{ss'\sigma}$ and $Z_{max}$ as adjustable parameters, while
keeping $\epsilon^0_{s'}$ constant and
$\epsilon^0_{s'}=\epsilon^0_s$ for simplicity. The parameter
values, which give the most reasonable moment variation with
respect to the coordination number of atoms, are listed also in
Table II.

After considering the spillover of electrons,
Weissmann's group\cite{guevara} introduced an intersite Coulomb
term to describe the valence level shift,
\begin{equation}
  \Delta \epsilon_i = \sum\limits_j \frac{U}{1+U|R_i-R_j|} \Delta n_j \label{weiss}.
\end{equation}
Equation (\ref{weiss}) returns to the intraatomic Coulomb repulsion as $R_i=R_j$,
and represents a bare Coulomb interaction without any screening
when atom $j$ is far from atom $i$. However, considering that the
spillover of electrons on the surface
may happen in a length scale longer than an atomic radius,
we think the interaction of the spilled electron might be different
from the usual intraatomic electron-electron interaction,
and the screening could be very strong for metallic materials.
As shown in the film calculations, both the charge and potential variations
are localized on the very top layer. We thus assume, in another extreme of strong screening,
that the orbital shifts are localized only on the surface atoms of a cluster.
We adopt a term,
 \begin{equation}
 \Delta \epsilon_{i'}(n^{i'}_{s'}) = \chi n^{i'}_{s'} \ \ \ \ \ \ \ \ i' \in {\rm surface},
\end{equation}
for this surface valence orbital shift, i.e., electron spillover
only changes the potential right on the surface atom. This shift
of orbital energy affects all valence (3$d$, 4$s$, and 4$p$)
orbits and the empty orbits $s'$ itself [see Hamiltonian (1)].
Proportional parameter $\chi=2.3$ eV (Table II) is found to give
the best fit of the surface valence orbital shift, calculated in
present paper for clusters, to the core level shift obtained in
the film geometry by LSDA calculations (see subsection V.A below).

Structures of very small clusters ($n=2$ through 8)
are very open, with average coordination number less than 5.
Parametrization used above for the description
of surface effects may lead to larger deviations.
Results presented below are only
for clusters with $n \geq 9$.

Let us now describe the process of going to full self-consistency
in present calculations. First, by switching off the exchange,
correlation and the spin-orbit coupling, i.e., setting
$I=U=\xi$=0, we get a density matrix containing the effect of
electron transfer between atoms but without spin and orbital
polarization. Then $I$, $U$ and $\xi$ are turned on, and
iterations continue by adding a finite uniform diagonal spin
polarization, i.e.,
       $\Delta_{\sigma}$ for spin $\sigma$ up, $-\Delta_{\sigma}$ for spin $\sigma$
       down,
to the non-polarized density matrix elements of all $d$ orbits.
Self-consistency is achieved
by solving Hamiltonian equation (1) iteratively.
From the self-consistent density matrix,
spin magnetic moment $\mu_{spin}$
and orbital magnetic moment $\mu_{orb}$ are obtained from
a vector average over the atomic moments:
\begin{eqnarray}
    \vec{\mu}_{orb}
  = \frac{1}{n} \sum_i \vec{\mu}_{orb}^{i}
  = \frac{1}{n} \sum_i \sum_{L\sigma,L'\sigma'}
       n^i_{L\sigma,L'\sigma'} (\vec{L})_{LL'}\delta_{\sigma\sigma'},
                           \nonumber   \\
    \vec{\mu}_{spin}
  = \frac{1}{n} \sum_i \vec{\mu}_{spin}^{i}
  = \frac{1}{n} \sum_i \sum_{L\sigma,L'\sigma'}
        n^i_{L\sigma,L'\sigma'} (2\vec{S})_{\sigma\sigma'}\delta_{LL'}.
\end{eqnarray}
The total moment, $\vec{\mu}$, is then calculated as their vector sum.

Although our iterations start from a uniform distribution of small spin
polarization and vanishing orbital polarization, the physical spin and
orbital moments obtained by self-consistently solving Hamiltonian (1)
could be non-uniform and non-collinear. However, it is found that the
non-collinearity is very weak.

Although SOC could generate anisotropy, we did not search for the energy minimum over all
possible directions, but only compared calculations with spin along
three principal axes of the clusters. We found that the energy differences are typically
less than 0.005 eV/atom. The effect of SOC on moment values is almost negligible,
and the moment differences for three directions are less than 0.02 $\mu_B$/atom.
Results reported below are calculated by setting the spin along the largest
inertia axis.

\section{Importance of Orbital Polarization}

In this section, we first separately discuss the spin and orbital contributions
to the total moments in connection with previous theoretical treatments, then
compare our results with experimental data. We show that orbital moments are
enhanced by over an order of magnitude for surface atoms in nano-clusters, and
dominate the oscillations of measured magnetic moment versus cluster size.
In contrast to bulk crystals where theoretically calculated spin moments agree well
with the measured total moments, orbital polarization is crucial to reach agreements
with experiments in nano-clusters.

\subsection{Size dependence of spin moment }

Spin moments calculated by Hamiltonian (1) are found almost the
same as those obtained by switching off orbital correlation and
SOC, i.e., setting U=0, and $\xi=0$, but keeping Stoner parameter
$I=1.12$ eV unchanged. One can compare our spin moments with
previous theoretical results of both LSDA and TB calculations, all
of which did not consider orbital correlation and SOC and give
only spin moment. Comparison with first principles results serves
as a critical justification of our TB parameterization scheme.

Figure 3 compares our results with those given by Reddy et
al.\cite{reddy}, who, using first principles molecular orbital
theory, studied the magnetic properties of Ni clusters in most
detail. For Ni$_{n}$ with $n=9$ through 14, and $n=19$, our
$\mu_{spin}$'s are in perfect agreement with their results. Either
Reddy et al.'s or present spin moments drop rapidly for small
clusters from $n=9$ to 13, in accordance with experiments (Fig.
1), but both calculated values are obviously lower than
experimental results. However, disagreements exist for $n=$15
through 18, and $n=20$, where our $\mu_{spin}$'s are larger than
Reddy et al. by about 0.2 to 0.4 $\mu_B$/atom. In order to
understand the origin of this discrepancy, we checked the
structures used in both calculations, and show the coordination in
Table III. It is found that for $n=9$ through 14, and $n=19$, the
MIAL structure used in present calculation is exactly the same as
that used by Reddy et al, but for $n=15, 16, 17, 18$, and 20,
there is big difference. We conclude that all differences between
present results and Reddy et al. are not from the theoretical
methods, though one is a LSDA and the other is a TB calculation,
but from the structures used in two calculations.

First principles calculations have been made also by other authors
on a few high symmetry clusters (Fig. 3). Using linear combination
of atomic-molecular orbital approach within the density functional
formalism, Reuse et al.\cite{reuse} get a magnetic moment 0.62
$\mu_B$/atom for Ni$_{13}$. Also, using the spin-polarized
discrete variational X$\alpha$ method, Fujima et al. obtained the
magnetic moments 0.58 and 0.73 $\mu_B$/atom for Ni$_{19}$ and
Ni$_{55}$, respectively\cite{fujima}. All these results are in
good agreement with present ones, $\mu_{spin}=0.57$, 0.68 and 0.66
$\mu_B$/atom for $n=13$, 19, and 55.

We conclude at this point that with present TB formalism and
parametrization scheme,  the calculated spin moments are in very
good agreements with {\it ab initio} results within an error bar
about 0.1 $\mu_B$/atom, as long as the calculations are made for
the same geometrical structures. This gives us confidences to
extend our TB studies towards large clusters which are hard for
direct first principles calculations. It also shows that a good
geometrical optimization is crucial when a detailed comparison is
to be made with experiments, irrespective of theoretical
approaches used in the studies of cluster magnetism.

We also note that all first principles as well as our spin moments
are significantly lower than experimentally measured values in a
rather wide size range. Our spin moments do show a minimum right
at $n=13$, but it is not as sharp as the experimental one. Even
more unexpected is that, contrary to experimental finding, our
spin moment shows very smooth size variations without oscillations
when $n \geq 22$, in obvious disagreements with experiments where
two minima exist at $n=34$ and 56 (vertical line in Fig. 3). We
demonstrate below that these discrepancies can be resolved when
the total moments, instead of the spin moments only, are compared
with the experimental moment values.

Before going to a discussion about the orbital contribution given
in present formalism, it is worth to compare also our results of
spin moment with two other important TB calculations. Since
cluster magnetism has been studied by Alonso et
al.\cite{alonso,bouarab,rodriguez} using exactly the same MIAL
geometry, it is thus of special importance for us to compare with
their results, in order to evaluate the influences of different
approximations in the adopted Hamiltonian. Alonso et al. have been
able to make their spin magnetic moments significantly larger than
ours and other LSDA results (Fig. 3). In their calculations, an
orbital energy shift $\Omega_{il}$ was introduced, which depends
not only on atom (subscript $i$ ) but also on orbit (subscript
$l$). These parameters were adjusted to make the number of $s, p$,
and $d$ electrons on each atom equal to values pre-assigned
according to its coordination number through a linear
interpolation between the isolated atoms and atoms in bulk
crystals. While this procedure approximately accounted for the
local surface effect, it completely neglected the redistribution
of electrons between $s$ and $d$ orbits which may happen due to
the global volume confinement. Besides, the valence energies of
different $l$'s were shifted so differently (even with different
signs), which is hard to understand on a Coulomb mechanism. Thus
Alonso et al's result of larger $\mu_{spin}$, though closer to the
measured total moment, was more or less an artifact, depending on
the pre-assignment of the number of electrons. In addition,
compared with experiments, their results did not present the right
positions (vertical lines in Fig. 3) of moment minima too.

In another TB calculation, Guevera et al.\cite{guevara} studied
the ideally cut fcc  Ni clusters. They used a Hamiltonian similar
to the ours, i.e., adding $s'$ orbits to surface atoms to consider
the electron spillover. In most cases, our results agree well with
those of Guevera et al (Fig. 3). Again, the spin moments by
Guevara et al were obviously lower than experimental ones, similar
to present results and other LSDA calculations. Since Guevera et
at. did not consider a complete series of cluster size, it is not
possible to check the moment minima positions to compare with
experiments.

At this point, we come to a conclusion that
the measured magnetic moments can {\em not} be considered
as only the spin moments, even though the calculated spin moments
do show enhancements in small clusters.
The prominent differences between measured
and calculated moments have to find other explanations.

\subsection{Orbital moment and its size dependence}

In principle, both spin and orbital magnetic moment have
contribution to the total magnetic moment in any system, though
the orbital one has been mostly quenched in bulk crystals.
However, it is shown that at the low
dimension\cite{zhou1,komelj,nonas,guirado}, the orbital quenching
could be released, and orbital polarization appears prominently.
In order to compare spin and orbital contributions, we plot spin,
orbital and total moments together in Fig. 4, as the functions of
cluster size. For Ni$_{n}$ with 22 $\leq$ n $\leq$ 60, $\mu_{orb}$
varies between 0.3 and 0.6 $\mu_B$/atom. Compared with the bulk
value, $\mu_{orb}$ has been enhanced over an order of magnitude.
Decreasing further the cluster size, $\mu_{orb}$ increases even
more. For example, $\mu_{orb}$ of Ni$_{9}$ reaches about 1.0
$\mu_B$/atom. We note that variations in orbital moments versue
cluster size are much larger than in spin moments. As seen from
Fig. 4, the oscillations of total magnetic moment mainly come from
the orbital contribution, which generates several minima in this
range of cluster size. Even at Ni$_{13}$, where both spin and
orbital magnetic moments contribute to the sharp minimum, the
orbital contribution is still the dominant one.

Of course, orbital moments calculated above depend on the choice
of the correlation parameter $U$. Unfortunately, there is no
standard value for $U$, and it is also uncertain how much it
depends on the cluster size, and how it varies from one atom to
another even in one single cluster. To ensure that our physical
conclusions drawn below are not limited to this particular choice
of parameter, i.e., $U=2.6$ eV, calculations are made also with
$U=1.8$ and 3.2 eV, in order to cover a wider range of generally
accepted values of correlation. Since the Stoner exchange is kept
as a constant, calculated spin moments change only slightly (less
than 0.1 $\mu_B$/atom) when $U$ changes from 1.8 to 3.2 eV.
Results of orbital moments are shown in Fig. 5 for comparison. As
expected, larger $U$ gives stronger orbital polarization.
Approximately, for $n=22$ to 60, orbital moments are 0.2-0.3,
0.3-0.6, and 0.4-0.8 $\mu_B$/atom for $U=1.8, 2.6$, and 3.2 eV,
respectively. Among these three parameter values, $U=2.6$ eV gives
the best fit to experimental results over the entire range of
cluster size (see next section).

However, not all featured minima positions
coincide each other for all three parameter values.
Only four ones exhibit independence on the choice of $U$ parameter
within the range from 1.8 to 3.2 eV, namely,
the moment minima at or near $n=13, 28, 34$, and 56
(vertical dashed lines in Fig. 5).
We will discuss the physical origin in following sections.
The remaining three minima at $n=19$, 23, and 40 exist only for $U=2.6$ eV,
but may disappear upon the change of $U$ value.
We have no reason to expect them appearing definitely in experiments,
and do not consider them in our further discussions.

\subsection{General comparison with experiments}

The total magnetic moments are plotted in Fig. 6 for Ni clusters
with $n=9$ through 60, compared with the experimental results of
Apsel et al\cite{apsel} and Knickelbein\cite{knickelbein}. Our
calculations reproduce many features given by the two experiments,
and the magnetic moment enhancement over the bulk values (0.61
$\mu_B$).

In the region between $n=9$ and 28, our results are in good
agreements with Knickelbein's experimental results, with
differences typically less than 0.1 $\mu_{B}$/atom. Our results
show the same trend of size dependence as Apsel et al from $n=12$
to 20, though our total moments are smaller by about 0.2-0.4
$\mu_B$/atom. In particular, both calculated and the two
experimental results show the pronounced sharp minimum at
Ni$_{13}$, and the significant moment increase as the cluster size
decreases to below $n=13$.

For Ni$_{n}$ with 30 $\leq$ n $ \leq$ 38, our results are larger
(about 0.1 $\sim$ 0.2 $\mu_B$/atom) than those of Apsel et al., but show
correctly the minimum at Ni$_{34}$. For Ni$_{n}$ with 40 $\leq$ n $\leq$ 60, our results
are in excellent quantitative agreement with Apsel et al. within
$\pm 0.05 \mu_B$/atom. The moment variation in this region is very smooth,
with only one minimum at $n=56$, also in agreement with Apsel et al.
A minimum at $n=28$ exists in our calculations,
but is not observed experimentally.
Because there is a structure transformation from MIC to TIC
right at $n=27$ to 28 (Fig. 2),
the MIC structure used in present calculation is obviously
not far from other metastable TIC structures.
A further check of the stability of the MIC structure of Ni$_{28}$
has to be made to clarify this discrepancy.

As shown in previous sections, the spin moments given by either
{\it ab initio} or TB calculations are much smaller than
the experimentally measured total moment, and do not exhibit
the correct oscillatory size dependence. After considering orbital correlation
and SOC, our calculations have reproduced not only the general trend of moment's size
dependence including all minima located at Ni$_{13}$, Ni$_{34}$ and Ni$_{56}$,
but also the absolute moment values measured in experiments.
This agreement shows clearly that orbital contribution is
an indispensable part in nano-cluster
magnetism, because not only $\mu_{orb}$ has been greatly enhanced
and compares fairly with its spin counterpart, but also the
moment oscillations are mainly due to the orbital contribution.

\section{Surface Enhancement}

In this section, we show that not only the spin moment, but also
orbital moment are enhanced for the surface atoms of the clusters
due to the change of their local environment. While the spin
enhancement is, as well known from previous studies on the
magnetism of ultra-thin films, due to surface valence orbital
shift and the increase of $d$ band holes induced by this shift, an
enhancement for the orbital contribution has also been induced by
this increase of $d$ holes, and is found even stronger. This plays
an important role in determining the size dependence of the
measured total moment, especially in the range of cluster size
larger than 20 atoms.

\subsection{Dependence on coordination}

Although there are no experimental techniques to probe the local
magnetic moment distribution in a cluster, it is still very
interesting to analyze the contributions from different sites inside a
cluster. This will help us to understand the size dependences of spin
and orbital magnetic moments, and provide further insights on the
enhancements of $\mu_{spin}$ and $\mu_{orb}$ in nano-clusters. To
check the suggestion of Jensen
and Bennemann, we plot in Figs. 7a and 7b respectively the spin and orbital magnetic moments
of all atoms inside three clusters --- Ni$_{42}$, Ni$_{55}$ and Ni$_{57}$,
as the functions of their coordination numbers ($Z$). We chose these three clusters
because our calculations yield the best agreements with
experiments in that size region (Fig. 6). In addition, they cover a wide
range of sampling by containing atoms with $Z=4$ through 12, and
exhibiting oblate, spherical, as well as prolate ellipsoids (Fig.
2).

Consider first the atoms with $Z=12$. The symmetry of Ni$_{55}$ is very high, possessing
13 atoms with $Z=12$, split into 3 non-equivalent types after including SOC.
For these atoms, we find $\mu_{spin}$ $\approx$ 0.54
$\mu_B$/atom, and $\mu_{orb}$ $\approx$ 0.06 $\mu_B$/atom (Figs.
7a and 7b) showing orbital quenching similar to that in the bulk
fcc Ni crystal. Although the symmetries of Ni$_{42}$ and Ni$_{57}$ are different
from that of Ni$_{55}$, atoms with $Z=12$ show similar moment values since the $d$
orbits of Ni atoms are very local. The agreement with the corresponding bulk values
$\mu_{spin} \approx 0.55$ $\mu_{B}$ and $\mu_{orb} \approx 0.05$ $\mu_{B}$ \cite{thole}, is another
evidence to show that our Hamiltonian parameters, namely, the $U$, $I$,
and $\xi$ are chosen reasonably.

We next consider $Z=8$ atoms. There are 30 such atoms in
Ni$_{55}$, again split into 3 non-equivalent types after including
SOC. As shown in Figs. 7a and 7b, their spin and orbital moment
differ not too much from those of $Z=12$ atoms. It is also true
for the $Z=8$ atoms in Ni$_{42}$ and Ni$_{57}$.

Finally, there are 12 atoms with $Z=6$ in Ni$_{55}$,
which are sorted into 2 non-equivalent types after including SOC.
These atoms have $\mu_{spin} \approx 0.93$ $\mu_B$/atom
and $\mu_{orb} \approx 0.7-1.0$ $\mu_B$/atom,
both obviously enhanced from the bulk values
(Figs. 7a and 7b).

Lower symmetry clusters Ni$_{42}$ and Ni$_{57}$ contain more types
of atoms. Atoms with even the same coordination could be different
by relaxation of bond length or bond angle etc. Due to this change
of environments, sites with the same coordination number may have
different $\mu_{spin}$ or $\mu_{orb}$, and this difference could
be quite large in low symmetry clusters (Figs. 7a and 7b).
Nevertheless, as shown obviously in the figures, lower
coordination number always results in larger local spin and
orbital magnetic moments.

For the sites with $Z \geq$ 8 (i.e. bulk and strongly bonding
surface atoms), both $\mu_{spin}$ and $\mu_{orb}$ are not enhanced
appreciably. On the contrary, for the sites with $Z \leq 6$ (i.e.
weakly bonding surface atoms), both $\mu_{spin}$ and $\mu_{orb}$
are greatly enhanced. This sharp distinction could be understood
from the general argument of Ref. \cite{zhou1}, that for a given
ratio between the orbital correlation and atomic bonding strength,
the ground state changes from an orbital polarized one to an
orbital quenched one at a certain dimensionality or coordination
number.

Although both local spin and orbital magnetic moments increase with
decreasing coordination, the enhancement of spin moment is smaller
than in orbital one. For example, while spin moment increase
from 0.55 $\mu_B$/atom for $Z=12$ to 0.7-1.2 $\mu_B$/atom for $Z=6$, orbital moment increases
from 0.05 $\mu_B$/atom to 0.4-1.2 $\mu_B$/atom, over an order of magnitude.
We note that the magnitude of orbital moment change is
0.35-1.15 $\mu_B$/atom, approximately twice as large as that for spin,
0.15-0.65 $\mu_B$/atom. This partly explains the stronger oscillations in $\mu_{orb}$.

As a numerical recipe, we found that the local spin and orbital
magnetic moments versus coordination can be approximated by :
\begin{equation}
\mu_{spin(orb)}(Z)=
\{
\begin{array}{ll}
\mu_{spin(orb)}^{bulk},& Z \geq 8 \\
\frac{1}{8}[(8-Z)\mu^{atom}_{spin(orb)}+Z\mu_{spin(orb)}^{bulk}],& Z < 8
\end{array}
\end{equation}
where, $\mu_{spin}^{atom}$ = 2.00 $\mu_{B}$
and    $\mu_{orb}^{atom}$  = 3.00 $\mu_{B}$,
taken from the Hund's rule values for Ni atoms,
and    $\mu_{spin}^{bulk}$ = 0.55 $\mu_{B}$
and    $\mu_{orb}^{bulk}$  = 0.05 $\mu_{B}$,
taken from the experimental values of bulk Ni crystals.
The interpolation results calculated by Eq. (11) are shown in Figs. 7a and 7b by dashed lines.
It is worth mentioning that $Z$-interpolation equation (11) holds approximately for all Ni$_n$
clusters with $n$ larger than 22, although it is derived from
Ni$_{42}$, Ni$_{55}$ and Ni$_{57}$ only.

Coordination deficit of surface atoms is known to have two
opposite effects to the spin magnetism, i.e., the
band narrowing which leads to a spin moment decrease for over
half filled 3$d$ metals, and the upward shift of surface valence
orbits due to the dipoles formed by the surface electronic spillover.
It is the latter effect which surpasses the former one, leading to an
overall giant surface moment, first revealed in early 1980s by LSDA
calculation\cite{freeman}. Figure 8 depicts the calculated surface
valence orbital shifts for Ni$_{42}$, Ni$_{55}$ and Ni$_{57}$
as the functions of the corresponding coordination numbers. It is shown that the
valence orbital shift is not sensitive to the cluster size, but
depends mainly on the coordination number. Because a site with
lower $Z$ has more {\it s$^{\prime}$} occupation, the shift
increases almost linearly with the local coordination deficit.

This valence orbital shift obtained in clusters by our TB calculations
could be compared with the core level shifts obtained in films by
standard LSDA calculations, because both of them are due to the
electron spillover and have the same physical origin. For
atoms on Ni (100) and (111) surfaces, their local coordination numbers
are 8 and 9, respectively. According to standard FLAPW calculations made for
5-layer slabs \cite{wimmer}, the corresponding 3$p_{3/2}$ core level shifts are
0.35 and 0.29 eV respectively (Table I and Fig. 8). They agree well
with those given by present TB Hamiltonian for nano-clusters. This
agreement shows that our parameterization scheme of the orbital
shift is more reasonable than that used by Alonso et al.
\cite{alonso,bouarab,rodriguez}, which required very different shifts for different orbits on even the same site.

\subsection{Coordination model and interpolation}

As shown above, local spin and orbital magnetic moments in larger clusters
depend mainly on local coordination numbers, being less sensitive to the cluster size.
Therefore, we expect that the $Z$-interpolation model, Eq. (11), can describe
the spin and orbital moment variations approximately.
As an extension of Jensen and Bennemann's coordination model \cite{jensen},
we assume that both local spin and orbital moments are determined
by their coordination numbers, with values taken from Eq. (11).
Neglecting the non-collinearity, the total moment of
a cluster is obtained by:
\begin{equation}
 \mu = \mu_{spin} + \mu_{orb}
     = \frac{1}{n} \sum\limits_{i} \mu_{orb}(Z_{i})
     + \frac{1}{n} \sum\limits_{i} \mu_{spin}(Z_{i})
\end{equation}

For clusters of the MIAL structure, $\mu_{spin}$ and $\mu_{orb}$
obtained from this $Z$-interpolation are shown in Fig. 9, together
with those calculated from the self-consistent TB model (already
shown in Fig. 4) for comparison.

For Ni$_{n}$ with $n \geq$ 22, the magnetic moments obtained from
this $Z$-interpolation agree well with those from our
self-consistent TB calculations. Their differences are less than $\pm
0.05$ $\mu_B$/atom for spin moments, and less than $\pm 0.1$
$\mu_B$/atom for orbital ones. This shows that the surface
enhancement effect given by Hamiltonian (1) could be reproduced
approximately by considering only the coordination dependence.

Since this scheme is so simple and elegant, it is of interest to
see to what extent it could be applied. Although there are experimental
measurements on magnetic moments for clusters with $n>$ 100,
no experimental and theoretical studies on geometrical structures are available.
We thus only consider nearly spherical clusters
cut ideally from a fcc crystal. In order to ensure these structures
to have the least relaxation on cluster surfaces which could be meaningfully
compared with experiments,  all clusters chosen to plot in Fig.
10 by the $Z$-interpolation possess atoms with coordination larger than
6.  The experimental results of Refs. 2 and 3
are shown in Fig. 10 as comparison. The $Z$-interpolation results are in good
agreements with those of Apsel et al. for clusters with $n<200$. When
$n>200$, the experimental moments are about 0.1 $\mu_B$/atom lower than the
$Z$-interpolation values. This difference might be due to the
temperature effect, because in this range of cluster size, Apsel
et al. made their measurement at an elevated temperature $T=303$ K.
As shown in their paper (Fig. 3 of Ref. \cite{apsel}), this elevation
of measuring temperature may indeed cause such an amount of moment decrease.
However, we can not offer a reasonable explanation to another
earlier experiment by Billas et al. which, though claimed to be performed
at temperature $T=120$ K, gave much lower magnetic
moments (Fig. 10).

We conclude at this point that for large Ni clusters ($n \geq$
22), local geometrical structure plays an important role in
determining the magnetic moments and a $Z$-interpolation model
gives good estimates on the size dependences of magnetic moment.
However, one should notice that the experimental moment minima
($n=13, 34$, and 56, shown by vertical lines in Fig.9), which are
quite prominent in TB calculations and independent of the choice
of $U$ parameter, become much less prominent in the
$Z$-interpolation results. Thus this $Z$-interpolation scheme is
not precise enough to give the delicate features on the size
dependence curve. Even more important point to be noticed here is
that, comparing with the TB calculation, for the small clusters
($n \leq$ 20), as shown in Fig. 9, the moments obtained by
considering only surface enhancement are rather large. For
example, at $n=13$, we have $\mu_{spin}=0.88 \mu_B$/atom and
$\mu_{orb}=0.73 \mu_B$/atom according to the $Z$-interpolation,
much larger than the corresponding TB values 0.57 $\mu_B$/atom and
0.15 $\mu_B$/atom. We will discuss the physics behind this huge
deviation in the next section.

\section{Quantum Confinement}

Since the moment variations in small clusters can not be explained
by surface enhancements only, we study the densities of states
(DOS) in such systems to explore other physics. The integrated DOS
per atom are shown in Fig. 11 for Ni$_{13}$, Ni$_{14}$ and
Ni$_{28}$. Enlarged plots are shown in Figs. 12a for an energy
window below the $d$ band bottom and in Fig. 12b for another
energy window near the Fermi (highest occupied) energy. It is
noted that different units of DOS are used in Fig. 11 and 12. It
is obvious from Fig. 11 that the electron occupation is dominated
by the $d$ bands, spreading from $-4$ to 2 eV with respect to the
Fermi energy. Due to the very localized nature of $d$ orbits, DOS
shown in Fig. 11 are very similar for all three clusters. However,
below the $d$ band bottom as shown in Fig. 12a, very
characteristic features due to quantum confinement appear. The
first two states, one spin up and another spin down, consist of
the atomic $s$ orbits of all atoms of clusters $n=13$, 14, or 28.
These two states show a global S symmetry, marked as 1S in Fig.
12a. The next six states (marked 1P in Fig. 12a) consist also of
the atomic $s$ orbits of all atoms inside a cluster, but have
global P symmetry over the whole cluster. Energies of these eight
states are lower than the $d$ band minimum, and are all occupied
for $n=13, 14$, and 28. Their energies decrease with increasing
the cluster size, and the degeneracy of 1P states depends on the
shape of clusters --- 1P states are perfectly degenerate for
nearly spherical clusters with $n=13$ and 28, but are splitted for
the prolate cluster Ni$_{14}$.

At a higher energy region, we find other ten states to consist
mainly of atomic $s$ orbits of all atoms but show a global D
symmetry. These states are marked as 1D in Fig. 12b, with five
states spin up and others spin down. For Ni$_{13}$, these ten
states are almost degenerate. The small energy differences are
induced by very small exchange interactions, because these states
contain mainly atomic $s$ orbits and only small $d$ components
through hybridization. For Ni$_{13}$, all ten D states are about 1
eV above the Fermi energy. For Ni$_{14}$, the structure is
prolonged along one direction and the symmetry is lowered. As a
consequence, the energies of two states, marked as 1D($\sigma$) in
Fig. 12b, are greatly lowered and become occupied; while the
remaining eight states, marked as 1D($\pi, \delta$) in Fig. 12b,
are still left empty above the Fermi energy.

This volume confinement leads to quantum terraces in the $s$
electron occupation against the cluster size as shown in Fig. 13.
With large enough size, we find the $s$-electron occupation to
increase almost linearly, about 0.72 $s$-electron per atom as
shown by the dashed line in Fig. 13. However, one terrace appears
from $n=9$ through 13 where the total number of $s$ electrons
remains 8 independent of the cluster size, as shown by the
horizontal solid line in Fig. 13. In fact, when the size increase
from $n=9$ through 13, all eight 1S and 1P states are occupied,
but all 1D states are empty due to the volume confinement.

The volume confinement on $s$ electrons has great impact
on the $d$ band filling. As the total number of electrons
increases from $n=9$ to 13, since the $s$ derived global S
and P states can hold only 8 electrons, the additional electrons
have to fill the $d$ derived bands. This is shown by the sharp
increase of the average $d$-electron occupation from $n=9$ to 13 in Fig. 14.
The $d$-electron occupation reaches a maximum at Ni$_{13}$.
Assuming full occupancy of the spin up $d$ bands,
the number of spin down holes could be simply estimated
by the quantum confined number of $s$ electrons,
giving 8/9=0.889, and 8/13 = 0.615 hole per atom,
for Ni$_9$ and Ni$_{13}$, respectively.
This is just the spin moments obtained
by all first principles LSDA calculations
(Reddy et al and Reuse et al in Fig. 3)
and approximately by TB calculations too
(present and Guevara et al in Fig. 3)
except that of Alsonso et al.
This explains why the spin moment of Ni$_{13}$
shows such a sharp minimum, and the total moment of Ni$_{13}$
drops even below that of many larger clusters in experiments.

The average $d$-electron occupation is
about 9.04 for clusters with $n > 30$ and is about
9.08 for clusters with $17 < n < 28$.
 Compared with Fig. 2, the step at 28 is probably related to
the structure transition. With increasing the cluster size, the
quantum confinement effect becomes small. Taking the difference between
TB and $Z$-interpolation result (Fig. 9) as a measure to describe this
quantum confinement, we note again that the effect on orbital moment is
roughly twice as that on spin moment.

Fujima and Yamaguchi \cite{fujima1} also considered this quantum confinement effect,
and assigned speculatively the oscillations of magnetic moment to the quantum filling
of a series of $s$ derived states.
Our calculations confirm that the minimum at Ni$_{13}$
is mostly due to this origin,
but other oscillations at larger size are mostly
from the surface enhancements of orbital moments.
Another point, different from their speculations,
is that even at just above Ni$_{13}$,
the filling of $s$ derived global D states does not happen abruptly
because the global D states are no longer degenerate
when a cluster becomes less symmetric when $n > 13$.
Instead, the ten global D states are lowered below the Fermi energy gradually,
leading to a gradual increase of spin moment when $n$ increases
from $n=14$ to about 20.

\section{Summary and Conclusions}

In addition to the hopping, exchange, and SOC terms which are
usually used in previous TB calculations, we show that orbital
correlation and valence orbital shift due to the electron
spillover from the surface atoms to the vacuum are important in
the study of nano-cluster magnetism. They have been included
properly in the TB Hamiltonian used in this paper. With this
rather general and unified formalism, we are able to show, in a
consistent way, that both spin and orbital magnetic moments of Ni
clusters are sensitive to the coordination numbers, and are
influenced by quantum confinement in very small clusters.

The calculated magnetic moments of Ni$_{n}$ clusters with $n$
between 9 and 60, are in better agreements with the experimental results
than previous theoretical treatments. For atomic sites with $Z \geq 8$
(i.e. bulk and strongly bonding surface atoms), both $\mu_{orb}$
and $\mu_{spin}$ are only slightly enhanced compared with bulk values,
but the enhancement is great for those with $Z \leq 6$ (i.e. weakly
bonding surface atoms). A simple coordination $Z$-interpolation is found to
account approximately for the size dependence of magnetic moment
when $n \geq $ 22. Thus, the coordination number (i.e. geometrical
structure) plays a main role in determining the magnetic moments of large Ni clusters,
and a proper TB calculation is necessary to give precisely the moment values and
exact positions of moment minima. Identification of minima is made
possible by analyzing our TB calculation results. Sharp minimum
at Ni$_{13}$ is contributed by both orbital and spin moments,
and due to both quantum confinement and surface enhancement effect. Minima at
Ni$_{34}$ and Ni$_{56}$ are contributed mostly by orbital
moments, and mainly due to surface enhancement effect.

Quantum confinement effect is the main reason that magnetic moment
of Ni$_{13}$ reaches such a pronounced sharp minimum. In small
clusters ($n \leq 20$), this effect has strong influences on the $d$
hole number, and $\mu_{spin}$ and $\mu_{orb}$ are greatly
decreased due to the decrease of $d$ holes. The change in
$\mu_{orb}$ due to decreasing $d$ holes is about twice as large as in
$\mu_{spin}$ for Ni atoms, as expected from Hund's rules for the
case when $d$ hole is less than 1. As a result, orbital
magnetic moment plays the most important role in the moment
oscillations versus clusters size.

In despite of the good agreements achieved, there are still some
discrepancies between experimental and our results. For example, a
moment minimum at $n=28$ appears in our theoretical calculations,
but it does not appear in experiments. This may be attributed to
the TIC to MIC structure transition at $n=27$, and 28. Also, it is
shown that for many other Ni clusters, energy differences between
the ground state and its closest isomer might be very small
\cite{xiang}. These might lead to discrepancies between
experiments and present theory, which is a subject of further
investigations.

{\bf Acknowledgments} XGW and DSW acknowledge the visiting scholar
program of National Center of Theoretical Sciences, Hsinchu,
Taiwan, which initiated this collaborative research. DSW also
acknowledge support by grants, No. 10234010, of National Science
Foundation of China, and No. G19990328-02, of National Pan-deng
Project of China. JMD and XGW acknowledges support by Grant No.
10074026 and Grant No. 10304007, of National Science Foundation of
China, and a grant for the State Key Program for Basic Research of
China.

\newpage

\begin{table}
\caption{Surface enhancement of spin moment
and core level (2$p_{3/2}$) shift of Ni films as calculated
from standard FLAPW method.}

\vskip5mm
\begin{center}
\begin{tabular}{lccc}
System              &  $Z$  & $\mu (\mu_B)$  & Shift (eV) \\
\hline
Bulk                & 12    & 0.561       &               \\
(111) film center   & 12    & 0.613       &               \\
(100) film center   & 12    & 0.619       &               \\
(111) film surface  & 9     & 0.625       &   0.291       \\
(100) film surface  & 8     & 0.675       &   0.354       \\
(111) monolayer     & 6     & 0.892       &               \\
(100) monolayer     & 4     & 1.014       &               \\
\end{tabular}
\end{center}

\end{table}

\newpage
\begin{table}
\caption{Parameters of Hamiltonian (1). Orbital energy, hopping integrals,
SOC strength, and Stoner exchange are taken from standard references.
Correlation parameter has been varied for comparison,
and parameters for surface empty orbit and the surface valence shift
have been chosen to give the most reasonable fit as stated in the text.
All values are in unit eV except the dimensionless $Z_{max}$. }

\vskip5mm
\begin{center}
\begin{tabular}{lcccccc}
Orbital energy & $\epsilon_{s}^{0}$ & $\epsilon_{p}^{0}$ & $\epsilon_{d}^{0}$
                           &&& \\
               & 15.50  & 24.64  & 9.74
                           &&& \\ \hline
Hopping integral ($\sigma$)
               & $V_{ss\sigma}$ & $V_{pp\sigma}$ & $V_{dd\sigma}$
               & $V_{sp\sigma}$ & $V_{sd\sigma}$ & $V_{pd\sigma}$
                                \\
               & -1.781 & 3.512 & -0.667 & 2.256 & -0.902 & -1.237
                               \\ \hline
Hopping integral ($\pi$)
               &                & $V_{pp\pi}$    & $V_{dd\pi}$
               &                &                & $V_{pd\pi}$
                               \\
               &                & 0.345          &  0.407
               &                &                & 0.266
                               \\ \hline
Hopping integral ($\delta$)
               &                &                & $V_{dd\delta}$
                           &&& \\
               &                &                & -0.037
                           &&& \\ \hline
SOC strength $\xi$    & 0.073         &&&&& \\ \hline
Stoner exchange $I$   & 1.12          &&&&& \\ \hline
Correlation $U$       & \multicolumn{5}{l}{2.6, and with options 1.8, and 3.2}
                             & \\ \hline
Surface empty orbit   & $\epsilon_{s'}^{0}$ & $V_{ss'\sigma}$  & $Z_{max}$
                           &&& \\
                      &  15.50              & -2.460           & 15.8
                           &&& \\ \hline
Surface valence shift $\chi$  &  2.3           &&&&& \\
\end{tabular}
\end{center}

\end{table}

\newpage
\begin{table}
\caption{Comparison between the structures used by Reddy et al and the MIAL structures
for clusters Ni$_n$ with $n=9$ through 20.
Listed are the number of atoms with coordination number $Z$ in the bracket,
and the number of total bonds. }

\vspace{5mm}
\begin{center}
\begin{tabular}{clclc}
n  & \multicolumn{2}{c}{Reddy et al} &\multicolumn{2}{c}{MIAL} \\ \hline
   &   atoms ($Z$)       & bonds & atoms               &bonds   \\  \hline
9  & 4(4),2(5),2(6),1(8) &  23   & 4(4),2(5),2(6),1(8) &  23  \\
\hline 10 & 3(4),3(5),3(6),1(9) &  27   &3(4),3(5),3(6),1(9)  & 27
\\ \hline 11 & 2(4),4(5),4(6),1(10)&  31 &2(4),4(5),4(6),1(10) &
31  \\ \hline 12 & 5(5),6(6),1(11)     & 36   & 5(5),6(6),1(11)
&  36  \\ \hline 13 & 12(6),1(12) &  42   & 12(6),1(12)         &
42  \\ \hline 14 &
1(3),9(6),3(7),1(12)&  45   & 1(3),9(6),3(7),1(12)&  45      \\
\hline 15 & 12(6),2(7),1(14)    &  50   &
2(4),8(6),2(7),2(8),1(12) &  49\\ \hline 16 & 1(4),7(5),7(6),1(9)
&  45   & 2(4),1(5),7(6),2(7),2(8),1(9),1(12) &  53\\ \hline 17 &
2(4),3(5),11(6),1(11)&  50  &2(4),2(5),6(6),2(7),3(8),1(10),1(12)
&  57\\ \hline 18 & 2(4),8(5),8(7)       &  52
&5(5),6(6),5(8),1(11),1(12) &  62\\ \hline 19 & 12(6),5(8),2(12) &
68  & 12(6),5(8),2(12)          &  68\\ \hline 20 &
2(5),16(6),2(11)     &  64  & 1(4),10(6),2(7),3(8),2(9),2(12) &
72\\
\end{tabular}
\end{center}

\end{table}

\newpage

\begin{figure}\caption{
Magnetic moments as a functions of cluster size of two experiments
(solid symbols), compared with first principles theoretical values
(open symbols). Four dashed lines show the experimental moment
minima positions.}
\end{figure}

\begin{figure}\caption{
Ratio of rotational inertia, $I_{1}/I_{3}$ and $I_{2}/I_{3}$,
and the evolution of the MIAL structures.
Four dashed lines show the experimental moment minima positions.}
\end{figure}

\begin{figure}\caption{
Spin moments of clusters with different sizes in present calculations
(square), showing good agreements with three first principles
calculations (open symbols) for most clusters, except for
$n=14,15,16,17,18,$ and 20, when different structures are used as
explained in the text. Two other TB calculation results (solid symbols)
are also plotted for comparison. Three vertical dashed lines show
the experimental moment minima positions.}
\end{figure}

\begin{figure}\caption{
Calculated spin, orbital, and total magnetic moments of Ni clusters.
Four vertical dashed lines show present theoretical moment minima positions.}
\end{figure}

\begin{figure}\caption{
Orbital moments calculated for correlation parameter
$U=1.80$, 2.60, and 3.20 eV.
Four vertical dashed lines show present theoretical moment minima positions,
which exist independent of the value of $U$ parameter in this range.}
\end{figure}

\begin{figure}\caption{
Comparison of calculated total moment of Ni clusters with the
experimental values by Apsel et al. and Knickelbein. Vertical
dashed lines show the theoretical moment minima positions.}
\end{figure}

\begin{figure}\caption{
Dependence of local spin (a), and orbital (b) moment on the
coordination number. Dashed lines represent an interpolation
between an isolated atom and that in a bulk crystal as given by
Eq. (11) in the text. Open circles denote the results of Ni films
calculated by the standard FLAPW method as given in Table I.}
\end{figure}

\begin{figure}\caption{
Valence orbital shift as a function of the coordination number,
compared with the core level shift (circles) of surface atoms from
standard FLAPW film calculations as given in Table I.}
\end{figure}

\begin{figure}\caption{
Comparison between the calculated spin and orbital moments by the
$Z$-interpolation model and TB calculations. Vertical dashed lines
show theoretically predicted positions of moment minima.}
\end{figure}

\begin{figure}\caption{
Total moment of clusters up to $n=700$, obtained by the $Z$-interpolation model
(crosses) and experiments (open sympols).}
\end{figure}

\begin{figure}\caption{
Integrated density of states per
{\em atom} for three clusters with $n=13,14$, and 28, with the Fermi level set as zero.}
\end{figure}

\begin{figure}\caption{
Integrated density of states per {\em cluster} in an energy window [-8 ev, -2 ev] for clusters
with $n=13,14$, and 28, and (b) in another energy window [-1 ev, 2 ev] for clusters with $n=13,14$.
Quantumly confined states are denoted in the figures by their glabal symmetries.}
\end{figure}

\begin{figure}\caption{
Total number of $s$ electrons as a
function of cluster size. Doted line shows the average number of
$s$ electrons per atom. Terrace (horizontal solid line) at total
$s$ electron number 8 from $n=9$ through 13 shows the quantum
effect of volume confinement.}
\end{figure}

\begin{figure}\caption{
Number of average $d$-electrons per atom as a function of cluster size.
The peak at $n=13$ explains the deep minimum of both spin and orbital moment
at this cluster size.}
\end{figure}

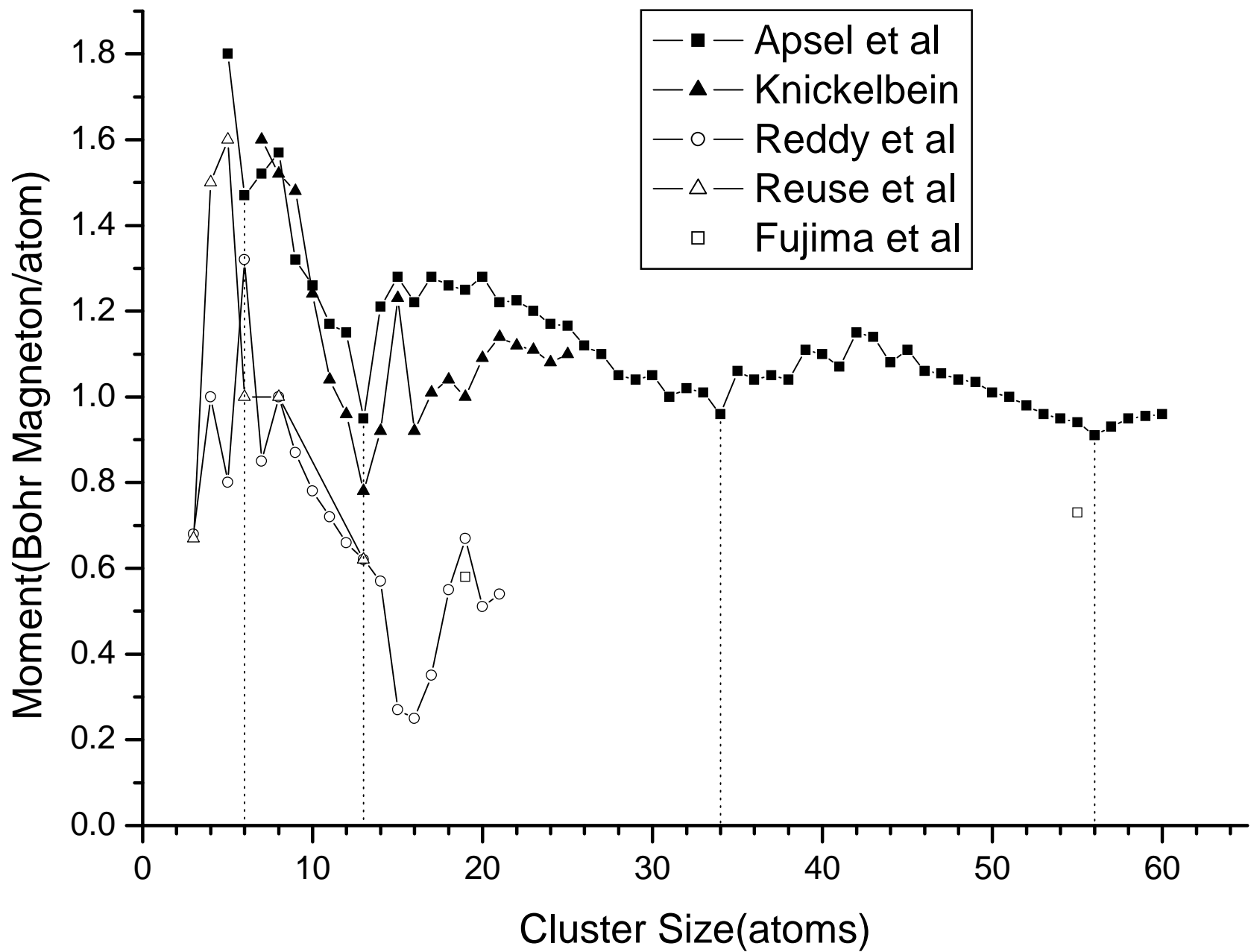
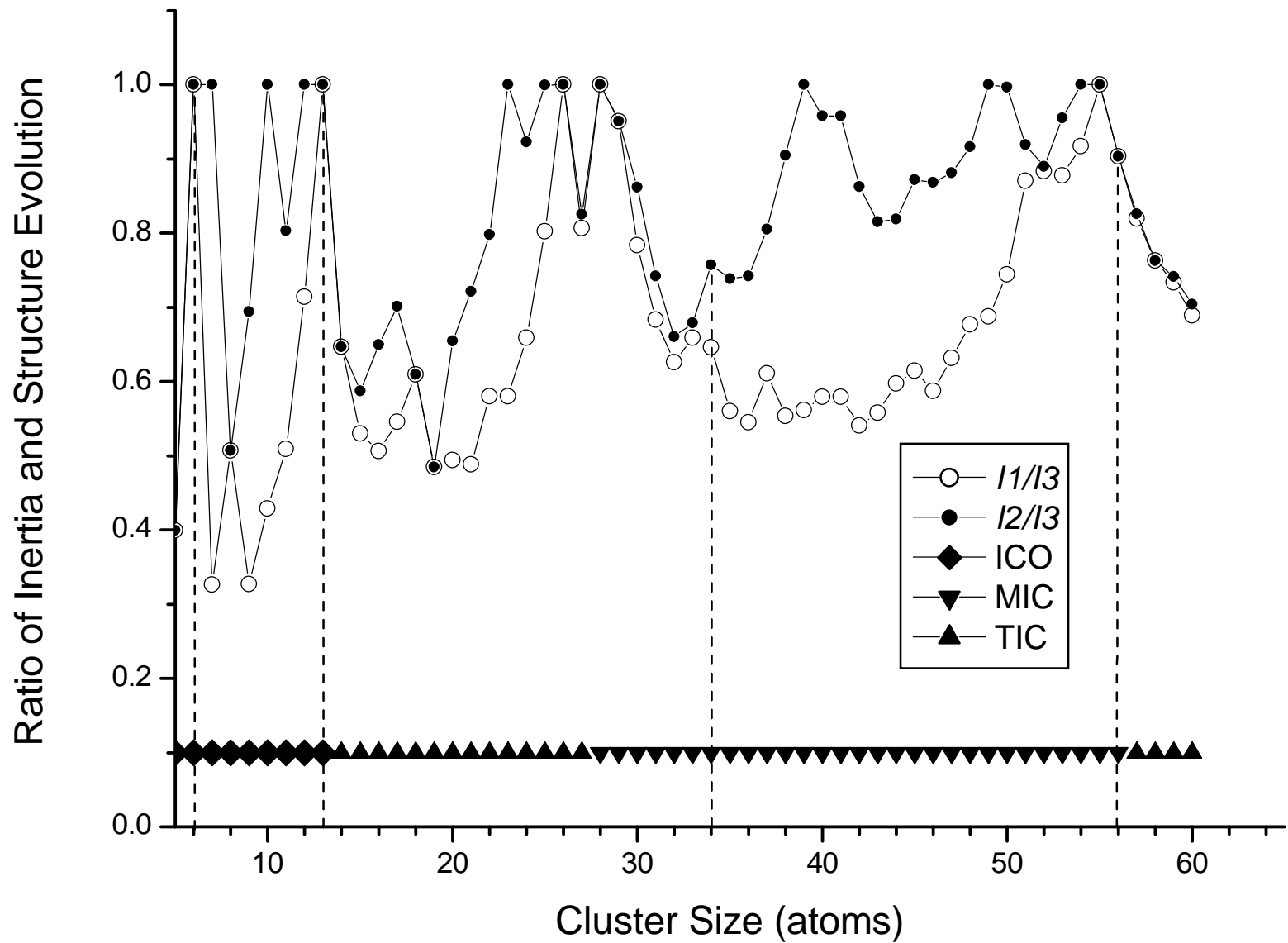
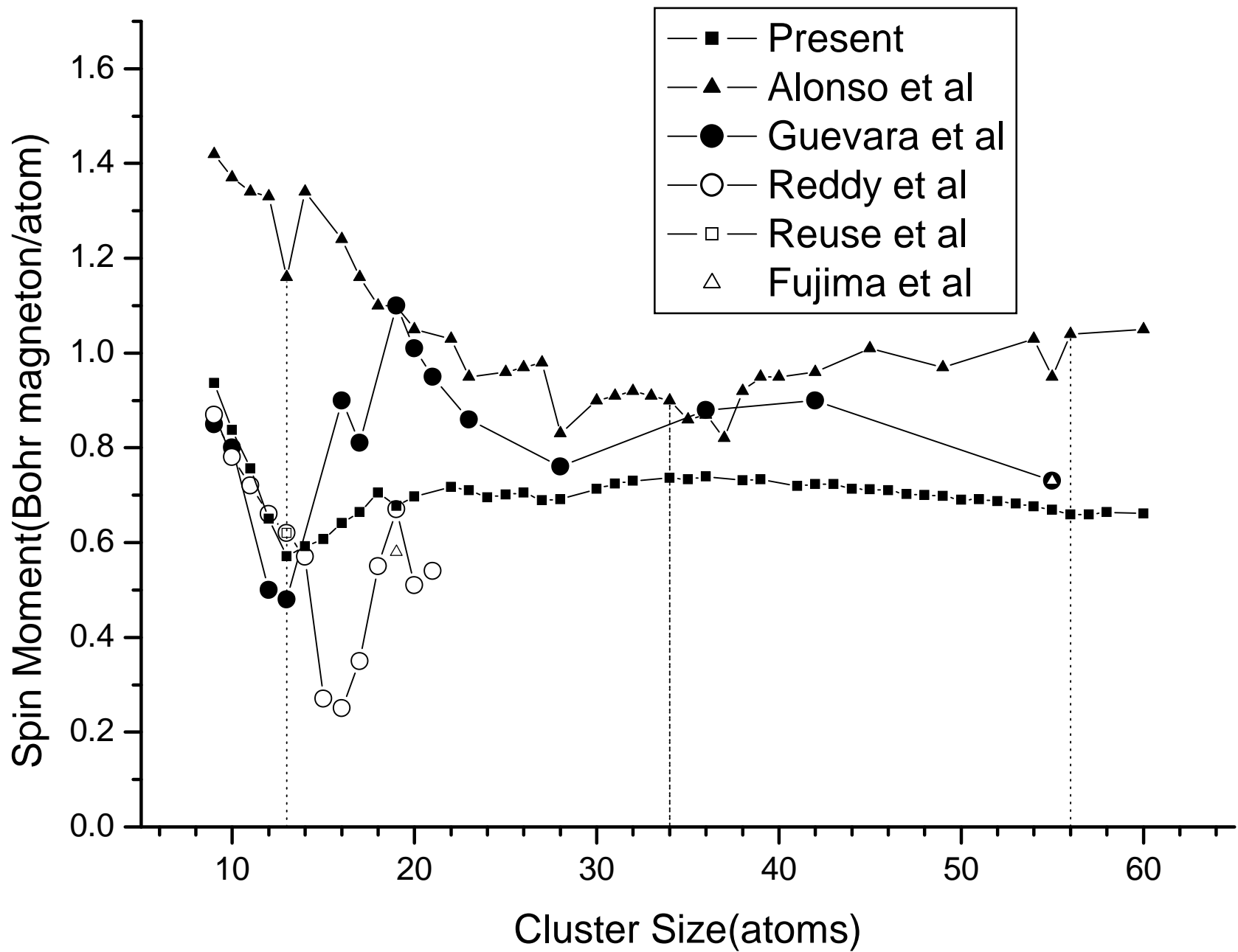
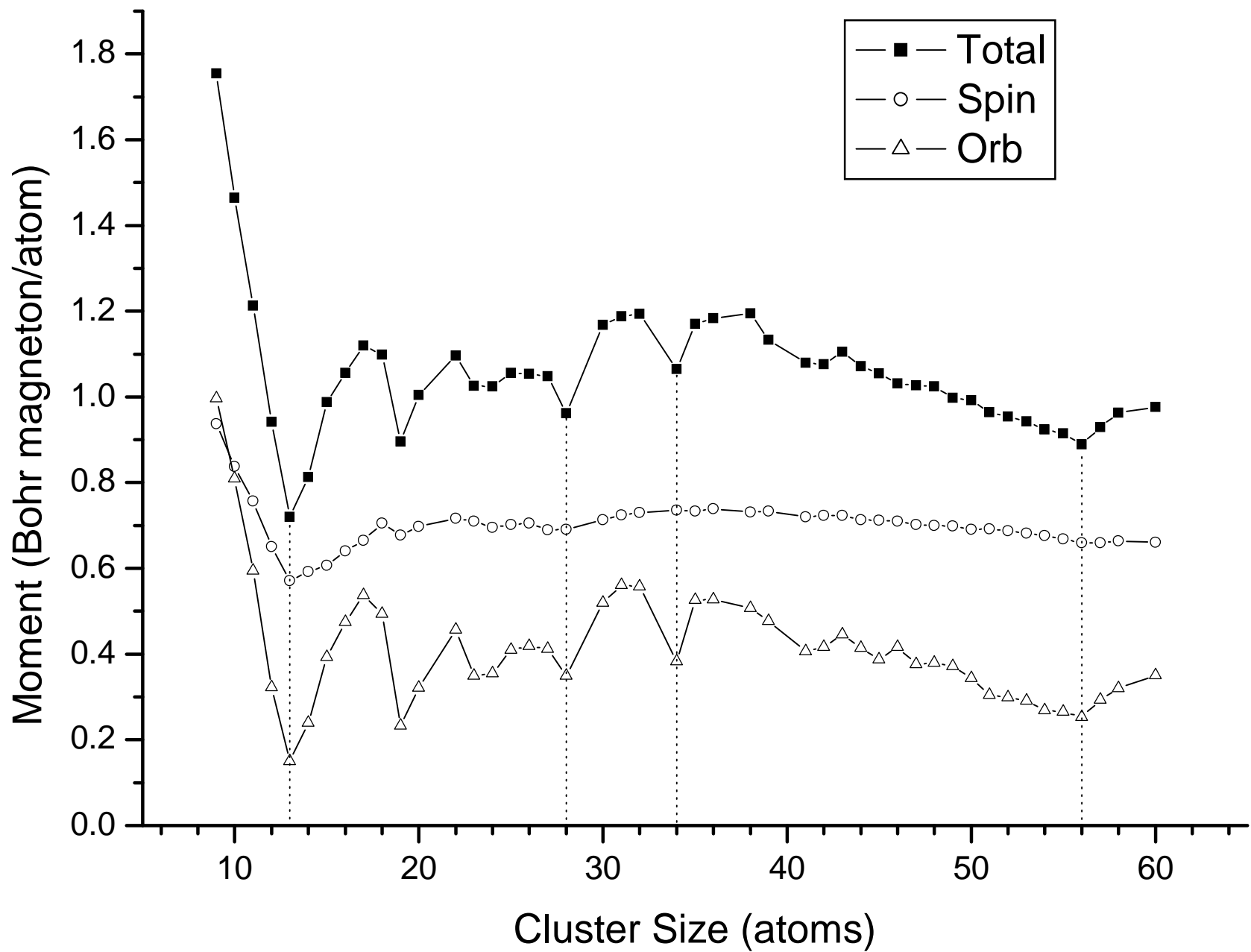
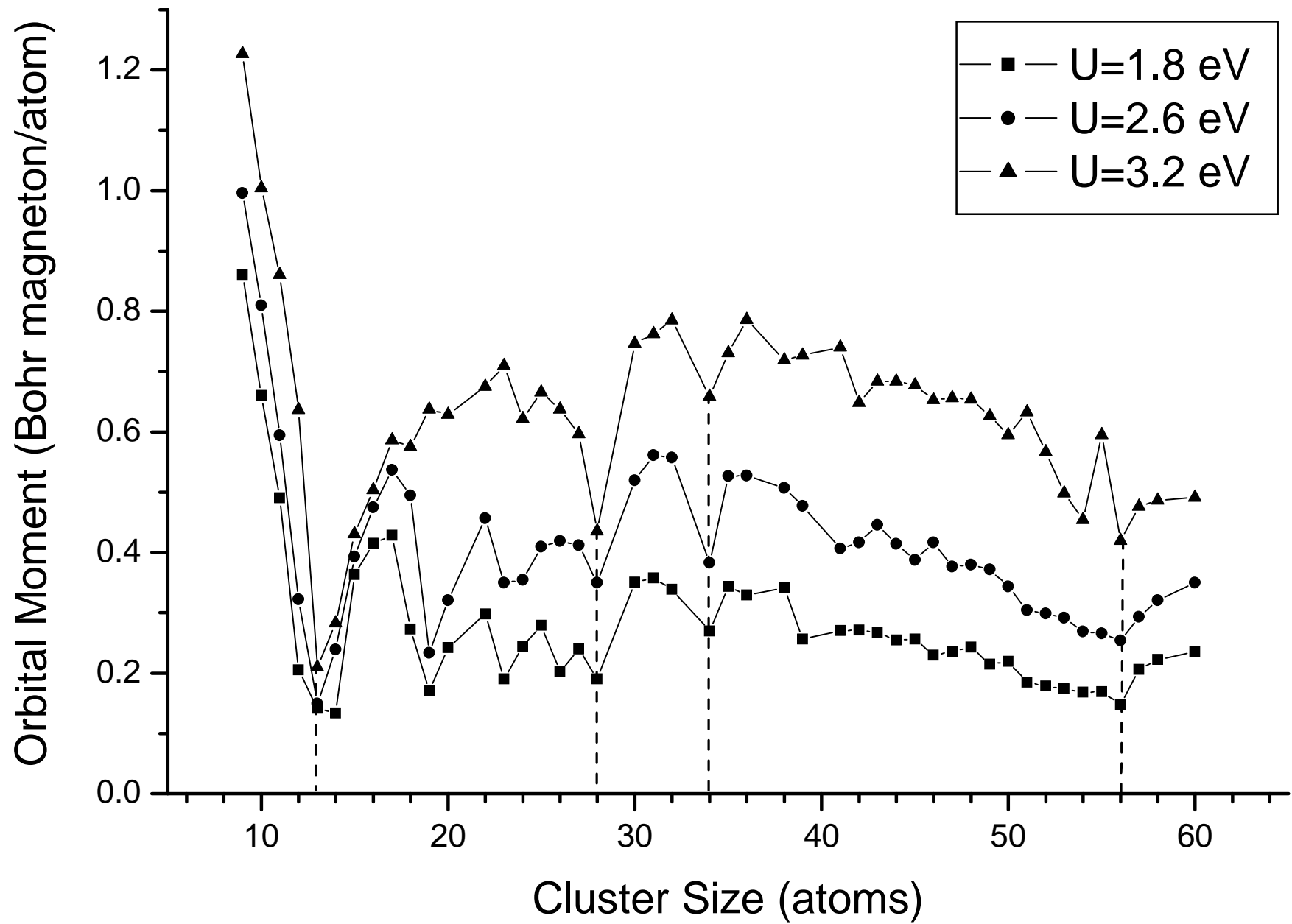
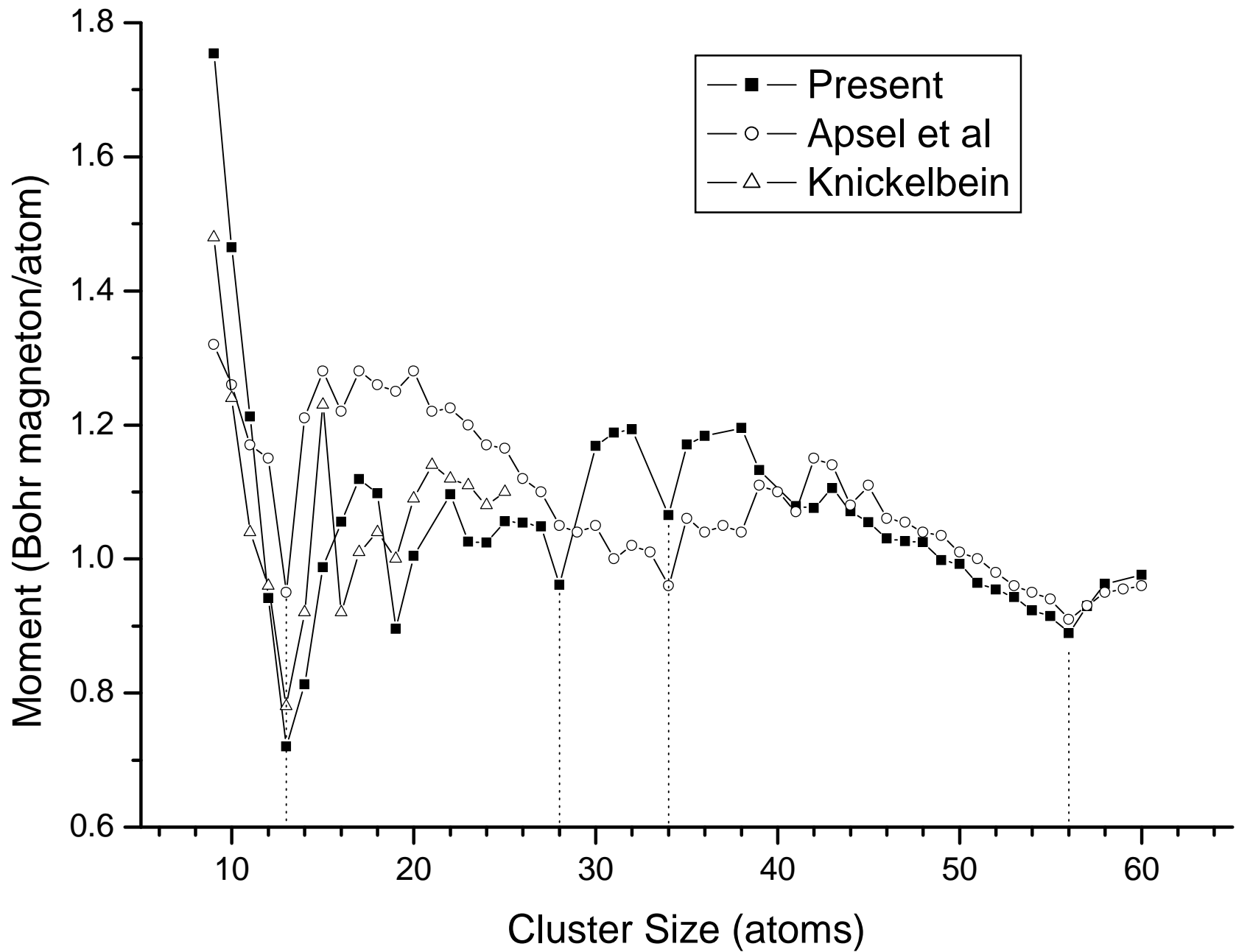
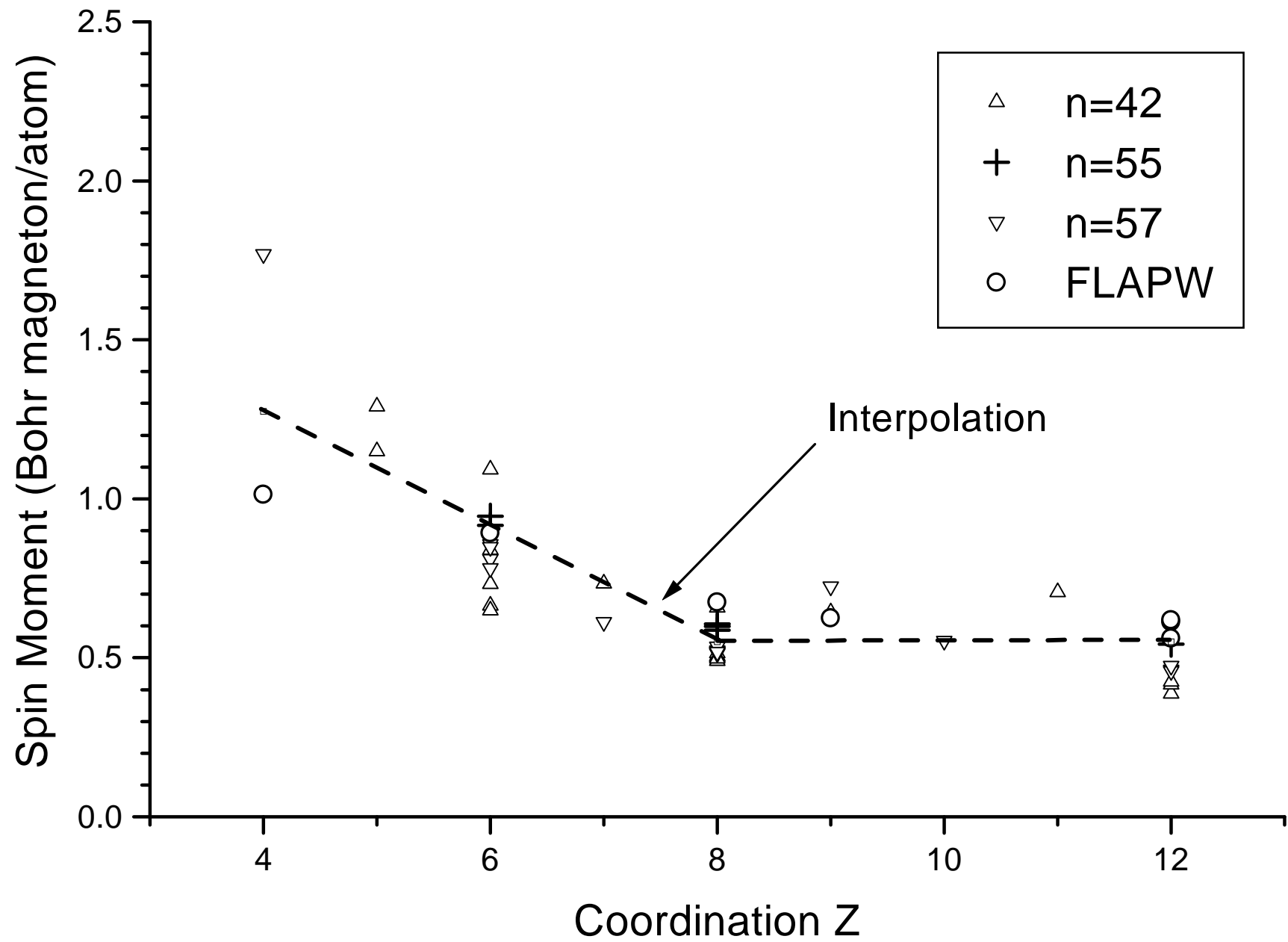
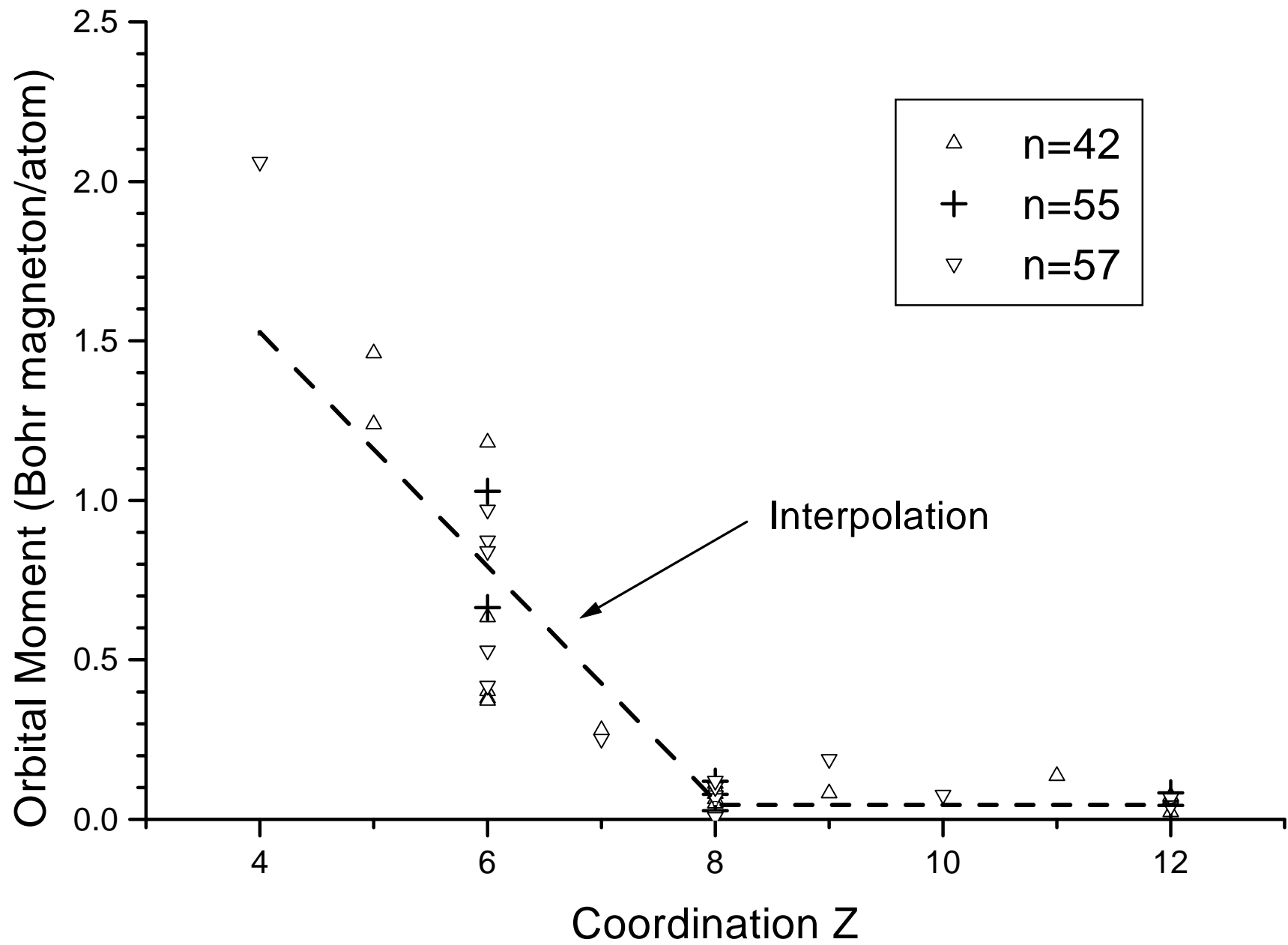
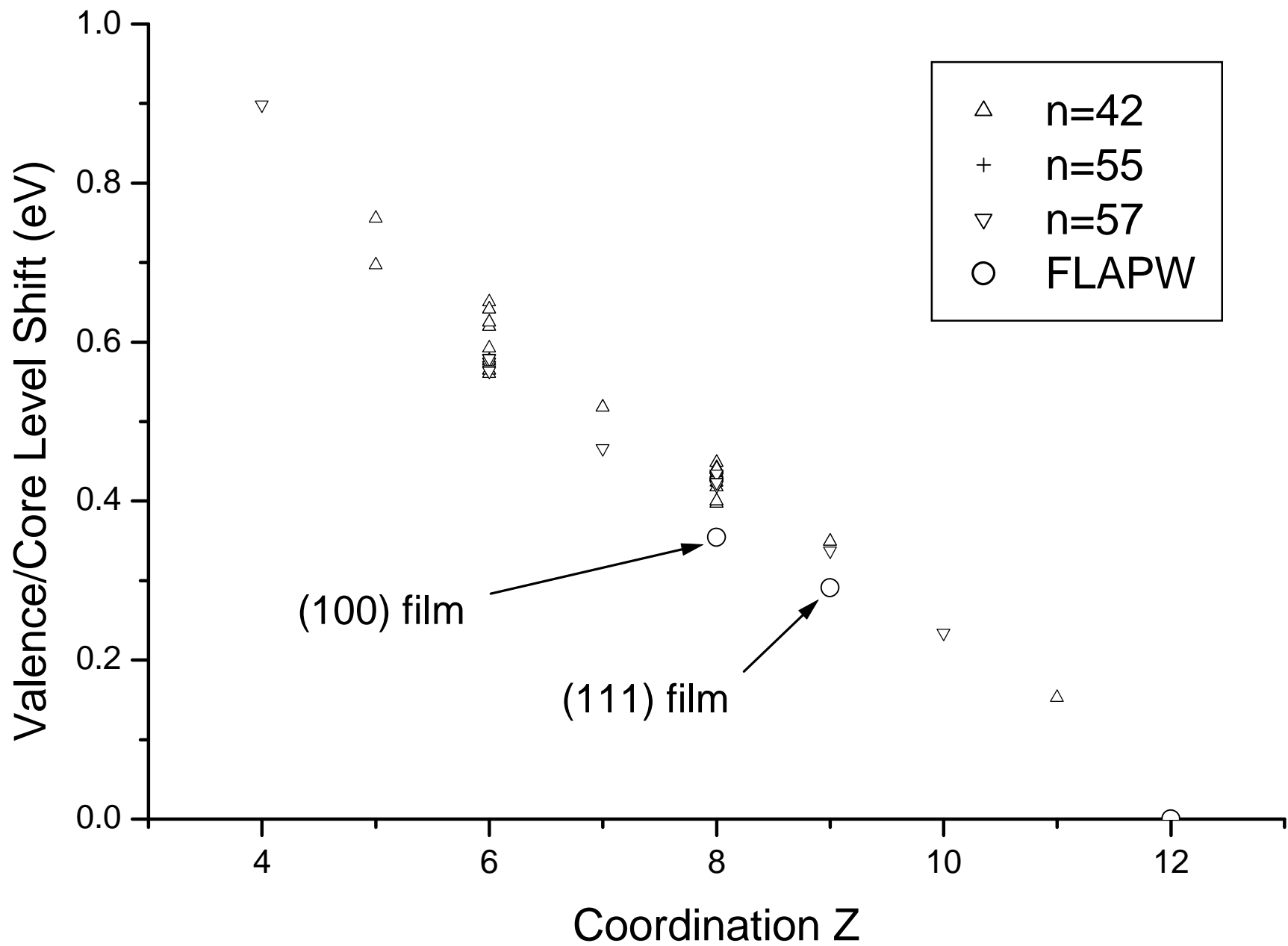
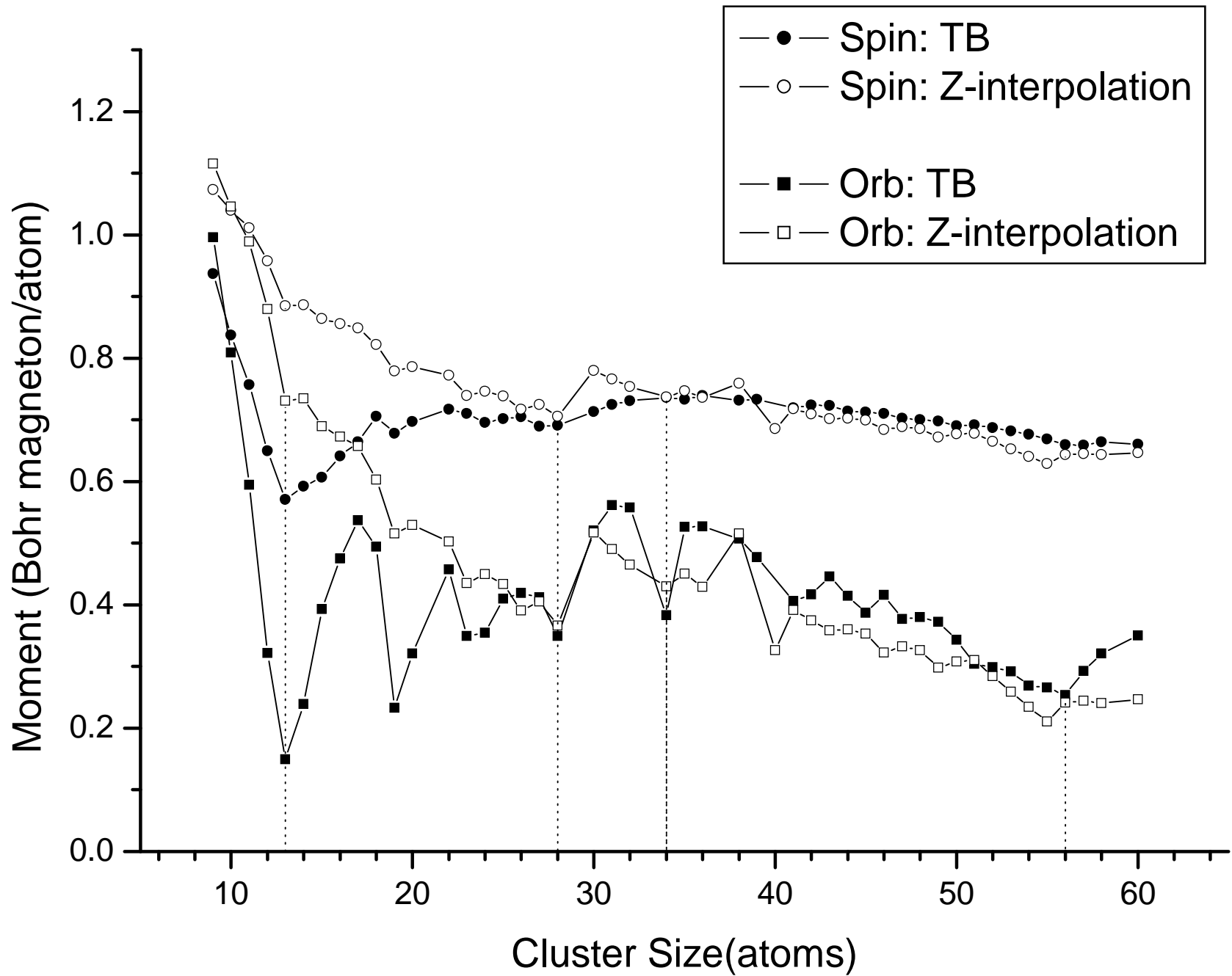
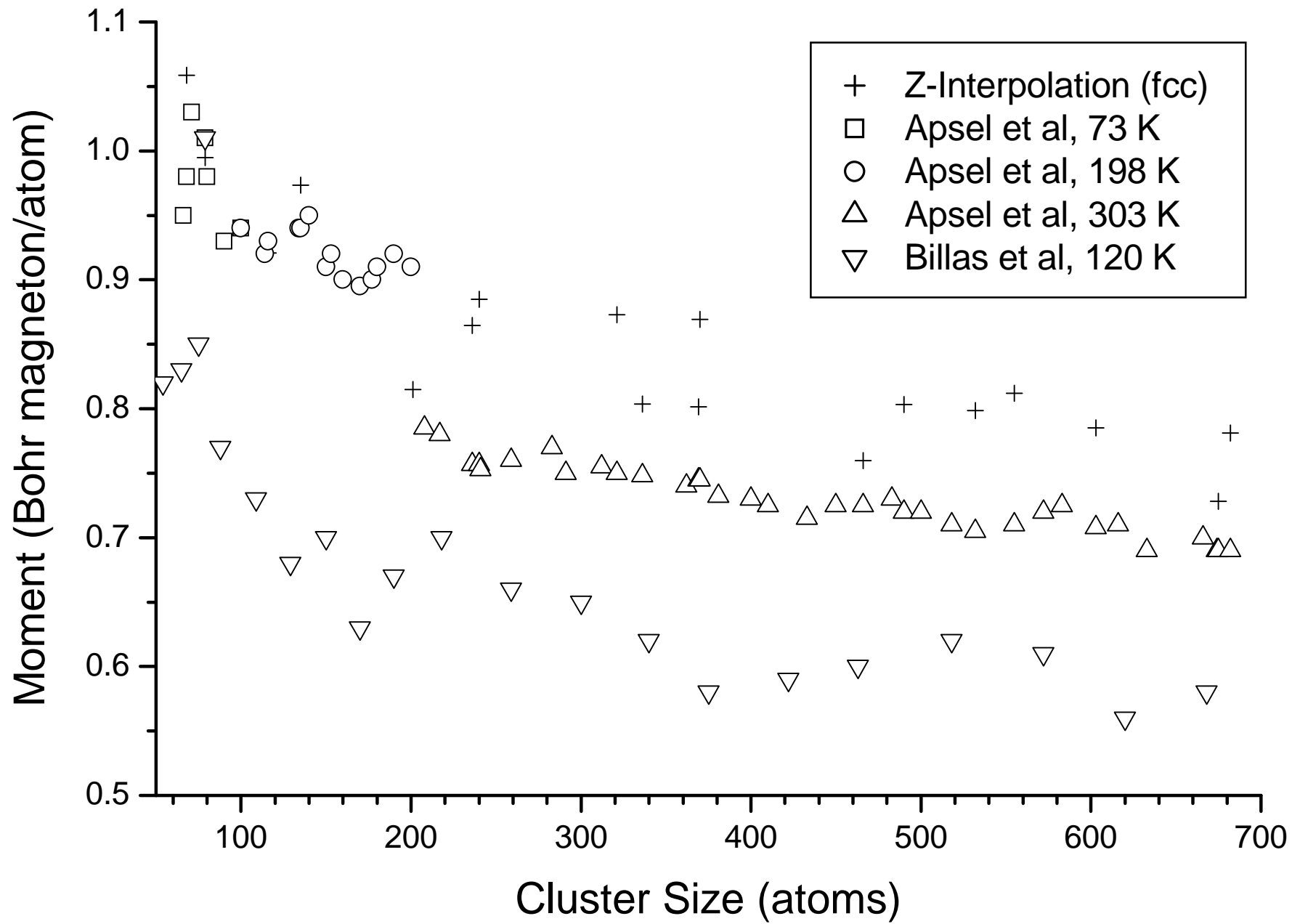
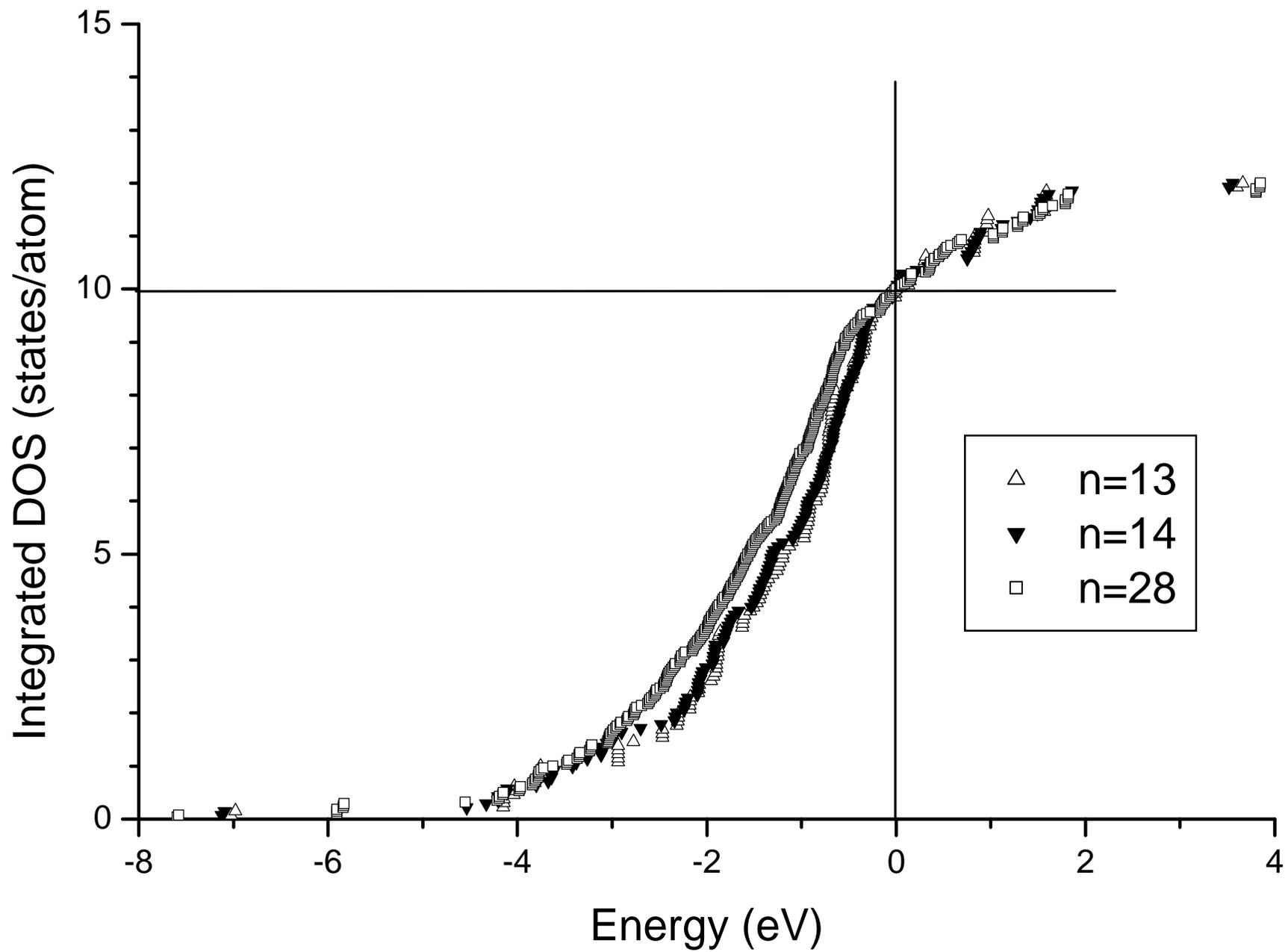
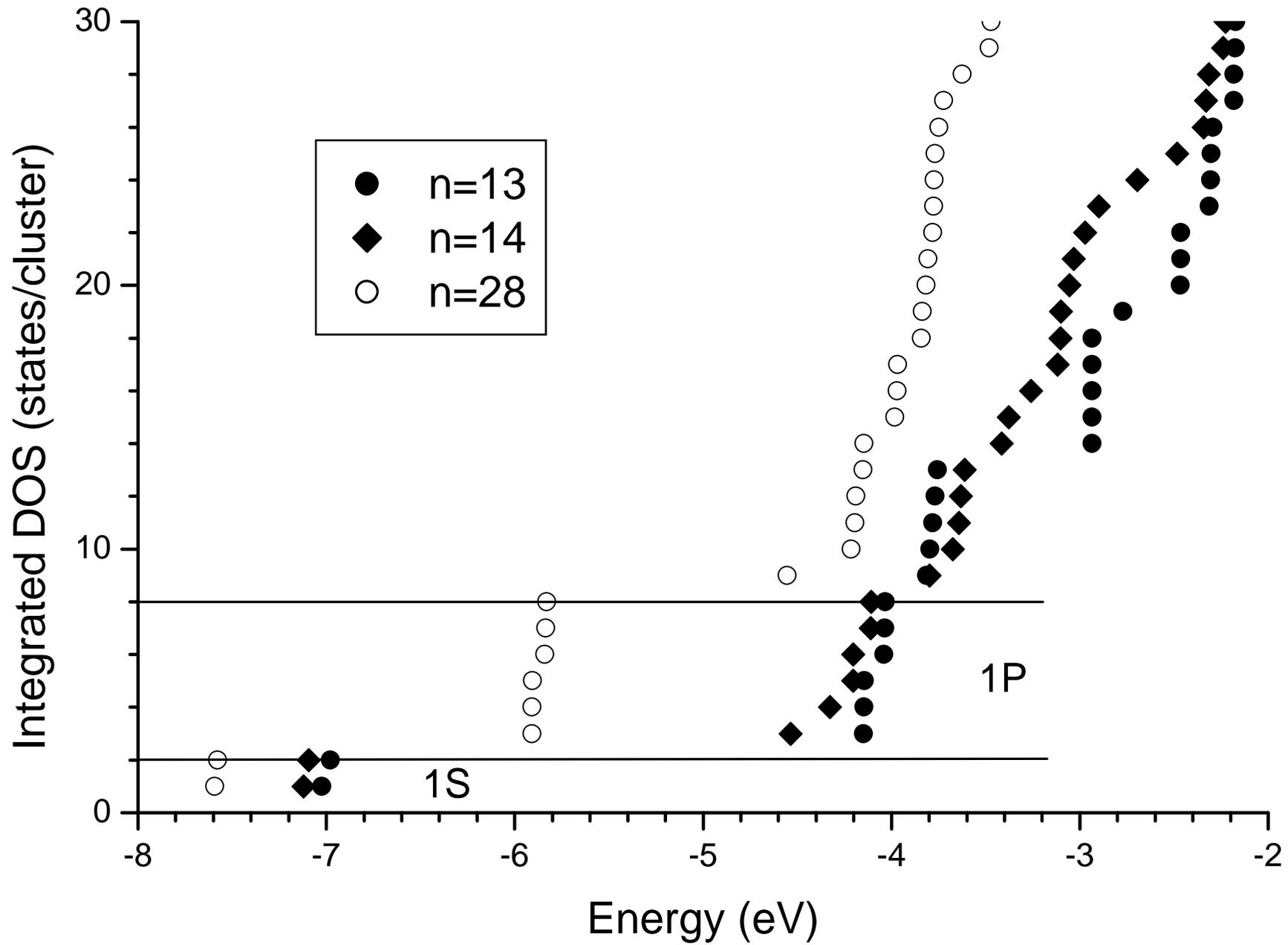
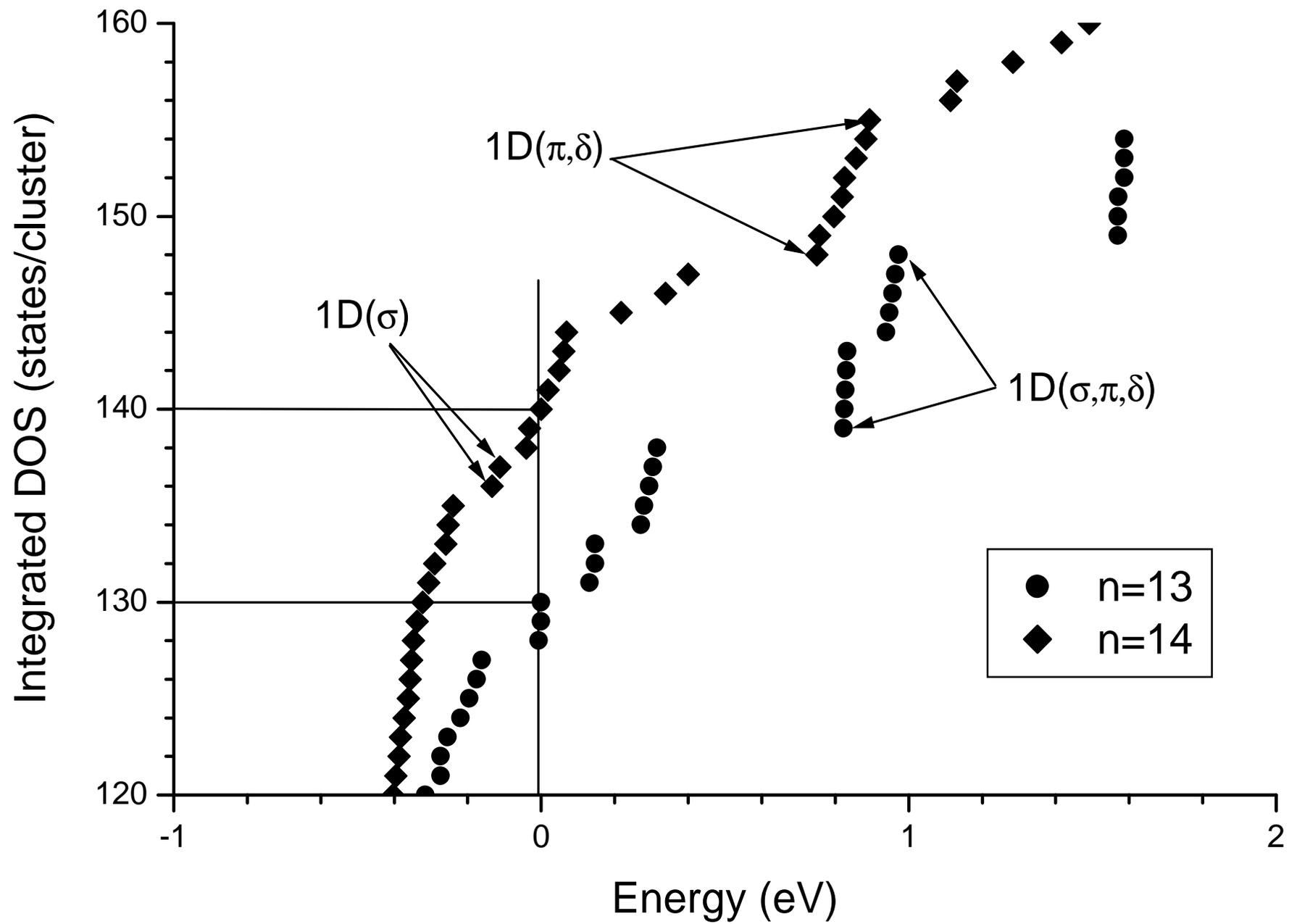
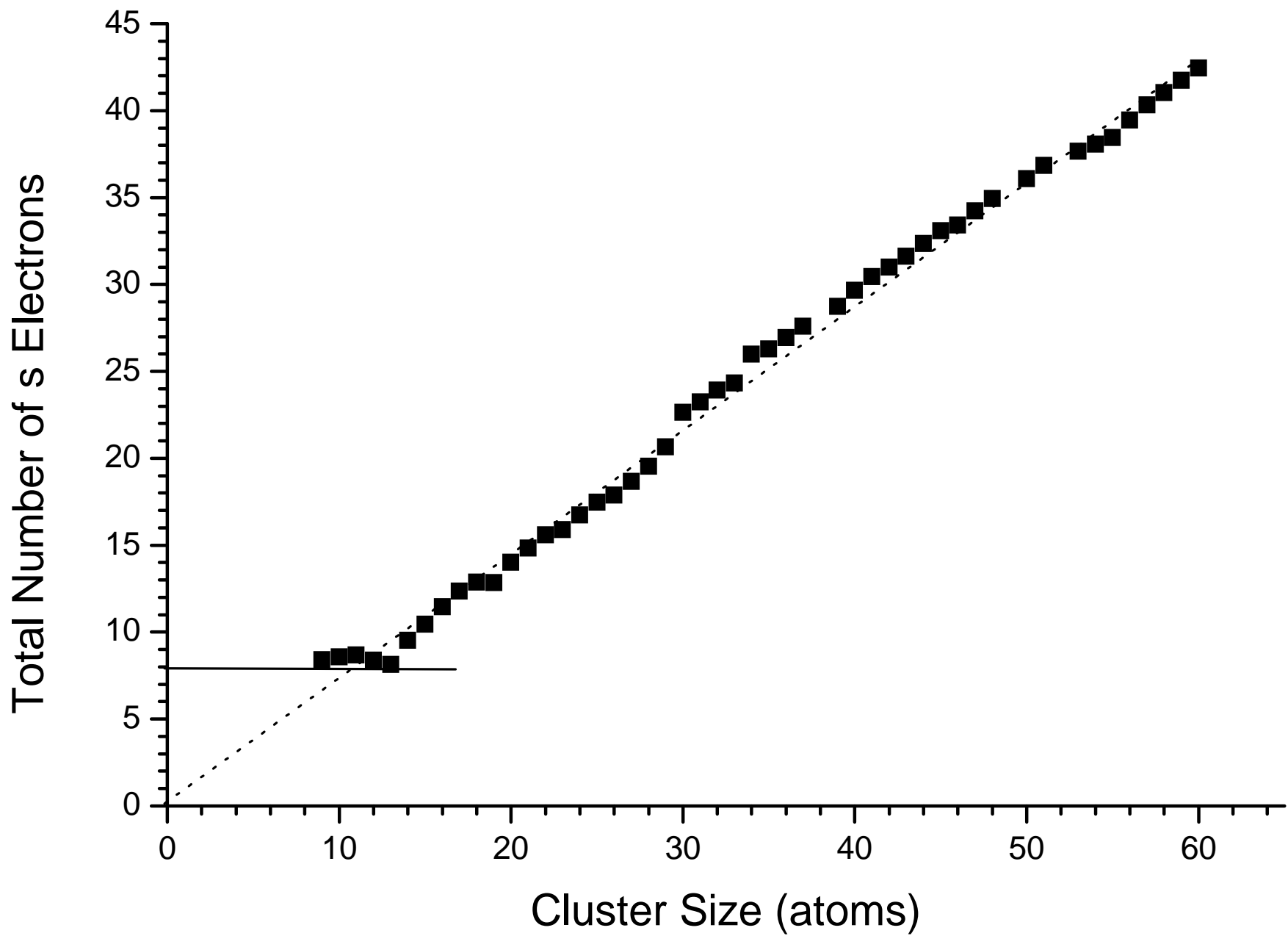
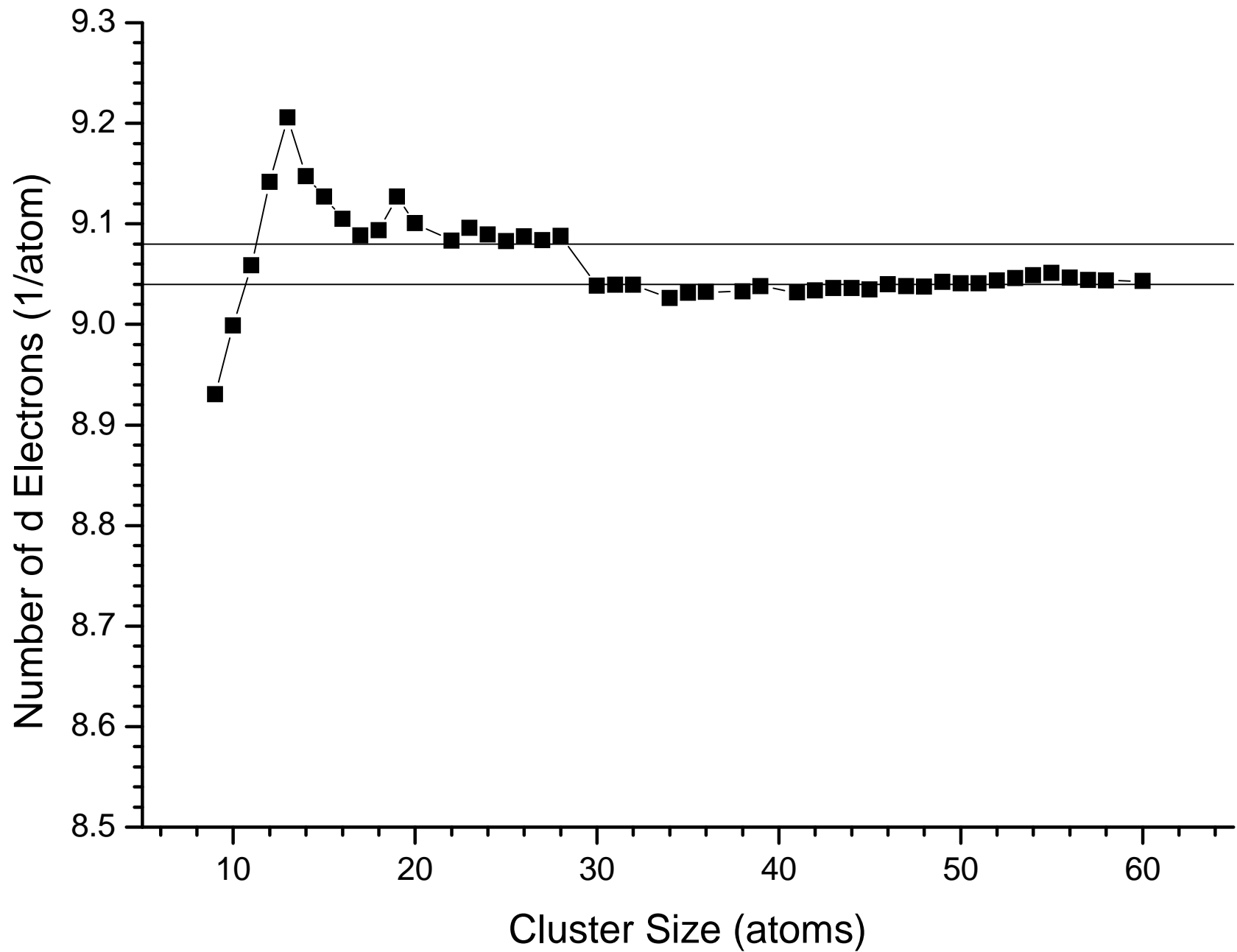
\end{document}